\documentclass[preprint,times]{elsarticle}
\pdfoutput = 1
\usepackage{amsmath}  
\usepackage{amssymb}
\usepackage{subfigure}
\usepackage{ecrc}
\usepackage{rotating}
\usepackage{xcolor}
\usepackage{latexsym}
\usepackage{bm}
\usepackage{floatrow}
\usepackage{breqn}
\newfloatcommand{capbtabbox}{table}[][0.45\textwidth]
\usepackage{tikz}

\volume{00}

\jid{}
\firstpage{1}
\journalname{}
\runauth{}

\begin{document}
\begin{frontmatter}

  \title{Yield criterion and finite strain behavior of random porous isotropic materials}

\author[CEA]{J.~Hure\corref{cor1}}

\cortext[cor1]{Corresponding author: jeremy.hure@cea.fr}
\address[CEA]{Universit\'e Paris-Saclay, CEA, Service d'\'Etude des Mat\'eriaux Irradi\'es, 91191, Gif-sur-Yvette, France}

\begin{abstract}
  The mechanical response of isotropic elastoplastic materials containing random distributions of initially spherical voids is investigated computationally based on Fast Fourier Transform simulations. Numerical limit-analysis simulations at constant stress triaxiality allow to determine the yield surfaces, leading in particular to the determination of a Representative Volume Element size for the onset of coalescence / inhomogeneous yielding. Moreover, two different coalescence regimes are observed that differ by the presence of shearing. The yield surfaces are found to be consistent with the combination of two models proposed in the literature, a GTN-type model calibrated for homogeneous yielding of random porous materials and an inhomogeneous yielding model accounting for both coalescence with or without shear. Finite strain simulations performed for different hardening moduli and stress triaxialities under axisymmetric loading conditions confirm the existence of a RVE up to the onset of inhomogeneous yielding. Coalescence strains are found to be significantly smaller for random porous materials than for periodic distribution of voids. A homogenized model is finally proposed that reproduces quantitatively the behavior of isotropic elastoplastic materials containing random distributions of voids under finite strains.
\end{abstract}

\begin{keyword}
Plasticity, Porous materials, Random void distribution, Inhomogeneous yielding 
\end{keyword}

\end{frontmatter}

\section{Introduction}

Predicting ductile fracture of metal alloys through void nucleation, growth and coalescence of internal voids requires modeling the mechanical behavior of porous materials. Physical mechanisms involved have long been recognized through early characterizations \cite{tipper,puttick}, and are nowadays described in details by dedicated model experiments \cite{weck,wecktoda} or large scale X-ray tomography experiments allowing to track voids evolutions \cite{yangshen,mairewithers}. Void growth \cite{ricetracey} and coalescence \cite{brown} phases are by far the most studied phases theoretically and numerically from a micromechanical perspective. Numerous homogenized yield criteria for porous metallic materials have been developed using different techniques, most of them using limit analysis \cite{gurson,thomason68} and more recently nonlinear variational approach \cite{danas}. Increasing complexity has been considered over the years, incorporating anisotropy - through Hill yield criterion \cite{benzergaanisotrope,keralavarma} or crystal plasticity \cite{xuhan,paux} -, void shapes effect \cite{gologanu,madou0}, or both \cite{morinellipse,mbiakop2}. These yield criteria have been validated through comparisons to porous unit cell simulations, and serve as key ingredients in homogenized models for porous materials that combine yield criteria and evolution laws for the additional state variables, the most well known being the so-called Gurson-Tvergaard-Needleman (GTN) model \cite{tvergaardneedleman}. The reader is referred to recent reviews about the various aspects of ductile fracture modeling \cite{benzergaleblond,besson2010,pineaureview,BLNT}.

Analytical homogenized yield criteria obtained through limit analysis are based on the definition of some simplified unit cell, \textit{e.g.}, spherical void in spherical cell \cite{gurson} or ellipsoidal void in a confocal cell \cite{gologanu}, that stands as a Representative Volume Element (RVE) in order to be able to perform the calculations. However, the effect of void distribution, \textcolor{black}{that can be observed either random \cite{mairewithers} or clustered \cite{HANNARD2017121}} from X-ray tomography experiments, has long been recognized as an important contribution to ductile fracture through void growth to coalescence. Early model experiments on perforated plates under tensile loading conditions have shown a lower ductility for random lattice of voids compared to regular arrays, as well as a dependence to the minimal intervoid distance \cite{dubensky1987,magnusen1988}. These effects have been assessed numerically by accounting for the inhomogeneity of void volume fraction by using the Gurson model \cite{becker1987}, or through extensions of Thomason's coalescence criterion \cite{melander1980}. The significant influence of the choice of the unit cell has been described in details for different periodic lattices in the void growth regime \cite{kunasun}, leading to heuristic corrections to the GTN model, and similarly in the void coalescence regime in the seminal contribution of Thomason \cite{thomason85a}. Progresses in terms of computational speed and numerical methods (\textit{e.g.,}with the use of Fast Fourier Transform (FFT) solver \cite{moulinec}) have lead to detailed descriptions regarding void distribution effect through three-dimensional finite strain unit cell simulations, based on synthetic \cite{thomson2003,tvergaard2017,VINCENT201474,WOJTACKI202099} or realistic void microstructures coming from X-ray tomography data \cite{shakoor2018,navas2018}. \textcolor{black}{The mechanical behavior of porous RVE has been assessed in \cite{bilger2005} considering either clustered or random void distributions.} The influence of both void number and void volume fraction have been investigated \cite{fritzen2012,fritzen2013,khdir2015,hoang2016}, leading to the characterization of the scatter associated with a given void number over different realizations. For low void volume fraction, the effect of void number is limited, justifying the use of single pore RVE as considered mostly in the literature. For large void volume fraction, a significant number of voids should be considered to obtain a RVE.

\begin{figure}[H]
  \centering
  \includegraphics[height = 4cm]{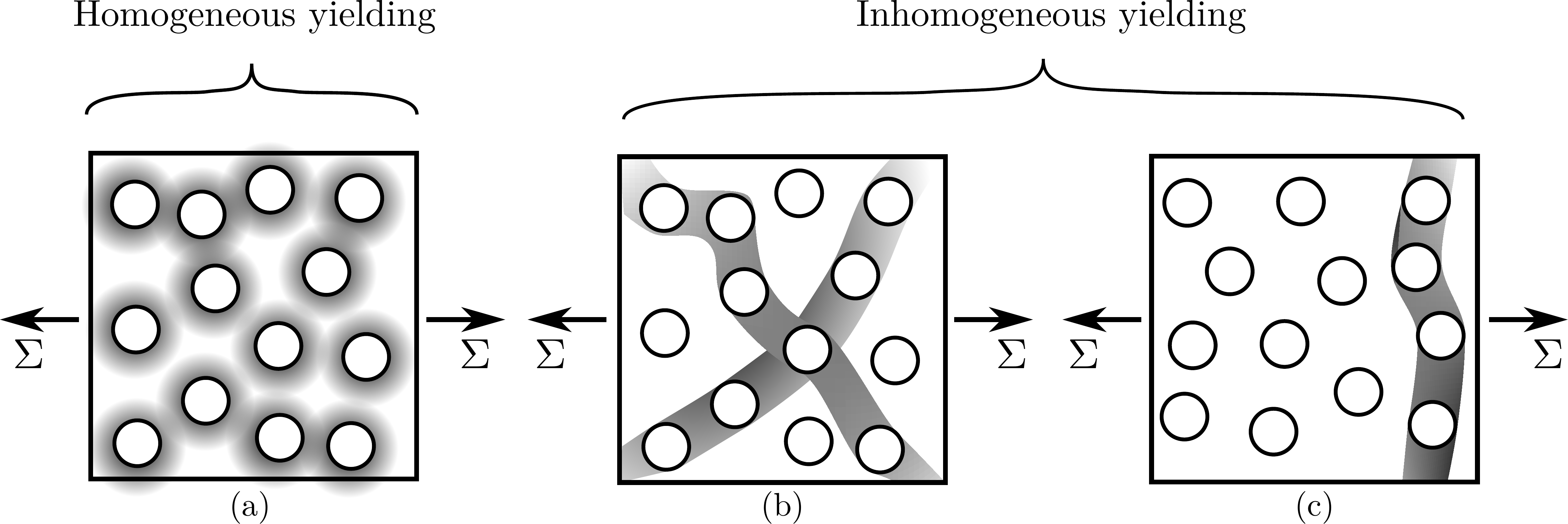}
\caption{\textcolor{black}{Local deformation mechanisms that can occurred in a random porous RVE: (a) homogeneous yielding and (b,c) inhomogeneous yielding}}
\label{fig0}
\end{figure}

\textcolor{black}{Homogenized yield criteria have been proposed for random porous isotropic materials, either through calibration of the GTN model parameters \cite{bilger2005,bilger2007,fritzen2012} based on random porous RVE simulation results or by using nonlinear variational approach (see, \textit{e.g.}, \cite{DANAS20122544,CAO2015240,song1}). These criteria assume that plastic flow is diffuse inside the RVE, and are referred to as homogeneous yielding conditions (Fig.~\ref{fig0}a). More recently, an homogenized yield criterion has been proposed for random porous isotropic materials \cite{KERALAVARMA2017100,REDDI2019190} considering inhomogeneous yielding (Fig.~\ref{fig0}b,c). This criterion builds on recent extensions of the Thomason coalescence model to account for shear with respect to the coalescence plane. Considering that plasticity occurs only in a layer composed of a regular array of voids of height equal to about one void's diameter, homogenized yield criteria have been derived theoretically in \cite{torki,keralavarma}. These criteria depend on the orientation of the so-called coalescence plane and have been applied to random porous materials in \cite{KERALAVARMA2017100} by considering an isotropic distribution of potential coalescence planes. The coalescence plane is thus selected only by mechanical loading conditions. Interestingly, the model can predict, for axisymmetric loading conditions, either uniaxial straining plastic flow (Fig.~\ref{fig0}c) or triaxial straining conditions if multiple local coalescence planes are selected (Fig.~\ref{fig0}b). In the latter case, the homogenized deformation mode appears as a \textit{pseudo-growth} mode that is very similar to the case depicted in Fig.~\ref{fig0}a, although the yield criterion is different. A multi-surface plasticity model for random porous plastic materials considering both homogeneous and inhomogeneous yielding  has been presented in \cite{VISHWAKARMA2019135}, including evolution laws for state variables, especially to account for hardening for inhomogeneous yielding.}

Up to date, the determination of yield surfaces for random porous materials through computational homogenization has been restricted to homogeneous yielding conditions \cite{fritzen2012}, or limited to few cases regarding void volume fraction for more general loading conditions \cite{bilger2005,VINCENT201474}. The effect of random void distribution and the occurrence of a RVE on coalescence / inhomogeneous yielding is still unknown, \textcolor{black}{and the associated homogenized yield criterion \cite{KERALAVARMA2017100,REDDI2019190} has not been compared to relevant - \textit{i.e.}, random porous materials - numerical data.} Therefore, the first objective of this study is to re-assess yield surfaces of random porous isotropic materials based on computational homogenization focusing on inhomogeneous yielding, and to assess the yield criteria proposed in the literature. Computational homogenization of random porous materials under finite strains remains scarce in the literature, mainly due to the associated numerical difficulties, but are required to validate ductile fracture models. Thus, the second objective of this study is to perform finite strain numerical simulations of isotropic materials containing random distributions of voids to assess a RVE size, and to compare to a homogenized model. \textcolor{black}{Only RVE composed of random distributions of voids are considered. The effects of void clusters will be discussed in the perspectives.}

The paper is organized as follows: in Section~\ref{num}, the numerical methods are presented. The determination of yield surfaces and RVE size through numerical limit-analysis are described in Section~\ref{ys}, as well as comparisons to yield criteria proposed in the literature. Section~\ref{fs} shows the results from finite strain computational homogenization and comparisons to the homogenized model predictions. Results obtained in this study are finally discussed in Section~\ref{disc}. 

\section{Numerical methods}
\label{num}
\subsection{Unit cells}

\begin{figure}[H]
  \centering
 \subfigure[]{\includegraphics[height = 3cm]{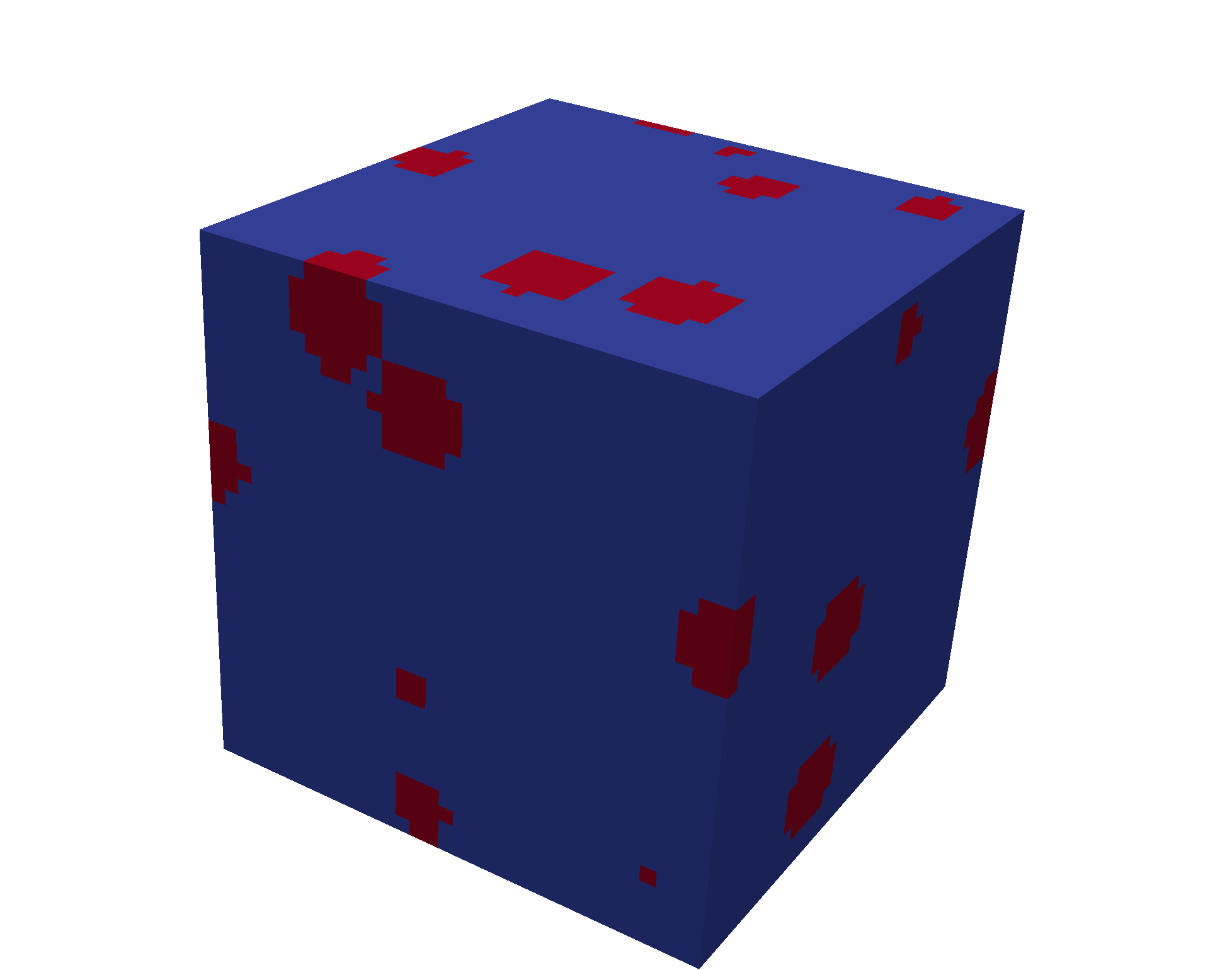}}
\subfigure[]{\includegraphics[height = 3cm]{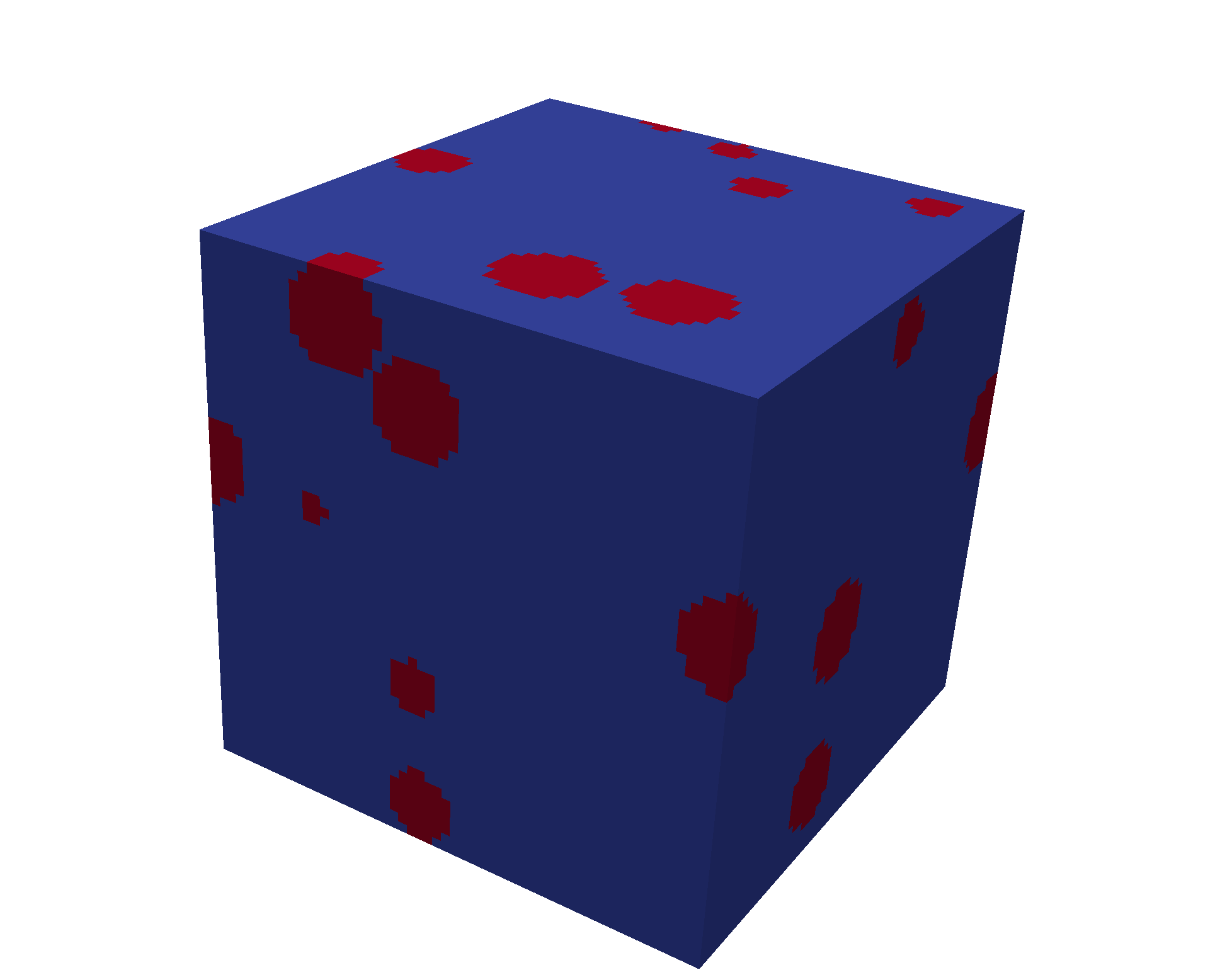}}
\subfigure[]{\includegraphics[height = 3cm]{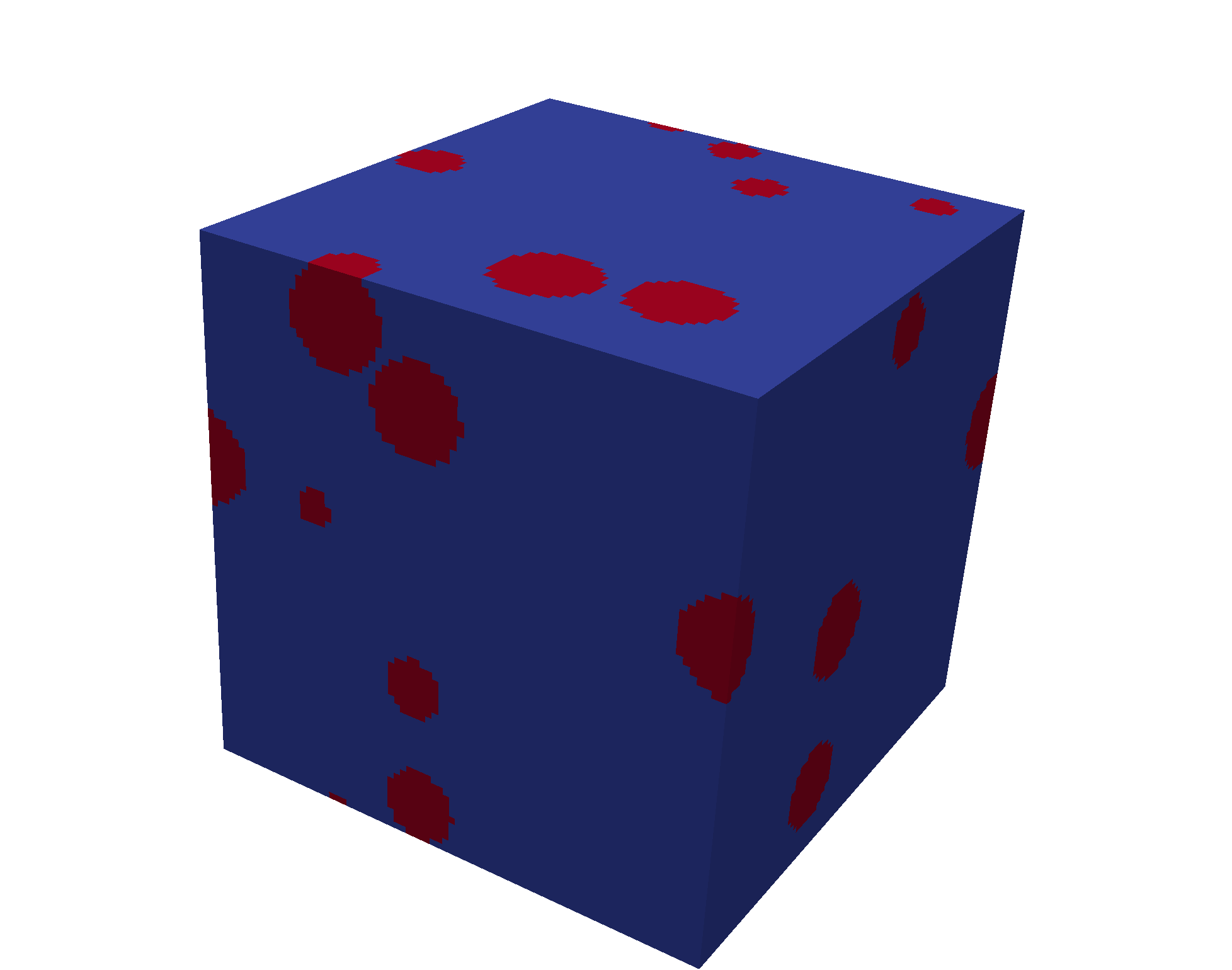}}
\caption{Examples of (pseudo-) random arrangement of voids (in red) used as porous unit cells in the following, for $N=32$ voids, porosity $f = 0.1$ and three different discretizations.}
\label{fig1}
\end{figure}

Cubic unit cells are considered in the following, composed of $N$ initially spherical voids of radius $R$. The position of the center of the voids are randomly chosen with the constraints of non-overlapping and ensuring the periodicity of the unit cell. Such void microstructures correspond to periodic hard-sphere models, as used for example in \cite{bilger2005,fritzen2012}\footnote{In \cite{fritzen2012}, as a minimal distance between two voids has been considered, the void microstructure is of cherry-pit type. The minimal distance value is however small, and used mainly for numerical reasons.}, and are in fact periodic clusters of $N$ voids, which describe truly random void distribution for $N \rightarrow +\infty$. $N=1$ corresponds to a cubic array of voids, which is widely used in porous materials modeling. Different seeds of the random number generator are selected to get different void distribution for a given void number. Regular grids are used to discretize the cells. For a particular discretization, the void size $R$ is adjusted such as to get the required porosity $f$. The choice to consider regular grids is dictated by the use of a Fast Fourier Transform (FFT) based solver, leading to stair-case void shapes, as voxels of the grids correspond either to void or to the sound material. However, it is shown in the following that this does not affect the homogenized behavior provided that the discretization is fine enough. Note that recent FFT simulations studies have proposed to use composite voxels \cite{CHARIERE20201} in order to avoid such discretization effect. Three typical examples of porous unit cells are shown on Fig.~\ref{fig1}, for $N = 32$ voids, a porosity of $f = 0.1$ and three levels of discretization.

\subsection{FFT based simulations and loading conditions}

Simulations are performed using the regular grids shown in Fig.~\ref{fig1} with the Fast Fourier Transform (FFT) solver \texttt{AMITEX\_FFTP} \cite{AMITEX}. Voxels corresponding to voids are stress-free, while voxels corresponding to the material follow von Mises plasticity with a Swift-type isotropic hardening:
\begin{equation}
  R(p) = \sigma_0\left(1 + \frac{p}{p_0}  \right)^m
\end{equation}
where $p$ is the accumulated von Mises equivalent plastic strain, $\sigma_0$ is the initial yield stress (taken as $\sigma_0 = 500\mathrm{MPa}$), $m$ the hardening exponent, and $p_0$ is set as $p_0 = \sigma_0/E$ with $E$ the Young's modulus (taken as $E = 200\mathrm{GPa}$). These constitutive equations have been implemented in the \texttt{MFront} code generator \cite{mfront}, using the Miehe-Apel-Lambrecht framework for finite strains \cite{MAL}. Periodic boundary conditions are considered, and macroscopic (volume-average) axisymmetric loading conditions are imposed:
\begin{equation}
  \underline{\sigma} = \begin{pmatrix}
    \sigma_{11} & 0 & 0 \\
    0          & \sigma_{22} & 0 \\
    0          & 0          & \sigma_{33}

  \end{pmatrix}
  = \sigma_{11}
  \begin{pmatrix}
    1 & 0 & 0 \\
    0          & \alpha & 0 \\
    0          & 0          & \alpha

  \end{pmatrix}
\end{equation}
All quantities (stress $\underline{\sigma}$ and deformation gradients $\underline{F}$) correspond to volume average quantities over the unit cell. All simulations are performed by imposing the axial deformation gradient $F_{11}$ in finite strain, or the deformation $\varepsilon_{11}$ in small strain. The parameter $\alpha$ is set to get the required stress triaxiality $T$:
\begin{equation}
  T = \frac{\sigma_m}{\sigma_{eq}} = \frac{1 + 2\alpha}{3(1-\alpha)}
\label{eqT}
\end{equation}
where $\sigma_m$ and $\sigma_{eq}$ are the mean stress and von Mises equivalent stress, respectively. These loading conditions correspond to one of the loading cases used in \cite{bilger2005}, and differ from those used in \cite{fritzen2012} (and subsequent studies by the same authors) as they do not prevent inhomogeneous deformation mode, which will be discussed in the following. Two kind of simulations are performed in this study, either in small strain - referred to as numerical limit-analysis - or finite strain settings.

\begin{figure}[H]
  \centering
  \subfigure[]{\includegraphics[height = 5cm]{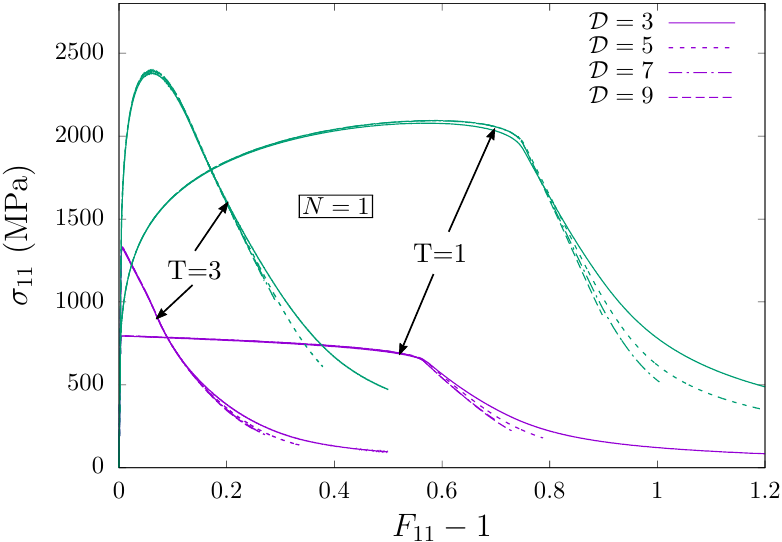}}
  \hspace{0.5cm}
\subfigure[]{\includegraphics[height = 5cm]{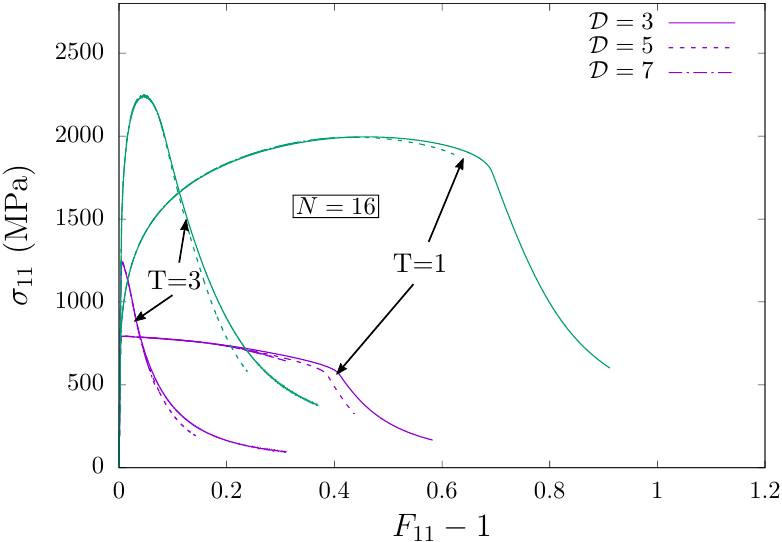}}
\caption{Evolution of the macroscopic Cauchy stress $\sigma_{11}$ as a function of the imposed macroscopic deformation gradient $F_{11}$ for porous unit cells with different hardening moduli $m \in [0.0;0.2]$, stress triaxiality ratio $T \in [1;3]$, discretization $\mathcal{D} \in [3;5;7;9]$  for (a) $N = 1$ and (b) $N = 16$ voids}
\label{fig2}
\end{figure}

The effect of discretization on porous unit cells FFT simulations results has been thoroughly studied in \cite{VINCENT201474,WOJTACKI202099} for small strains, leading to a number of voxels along each void radius between about 16 \cite{WOJTACKI202099} to about 32 \cite{VINCENT201474} to achieve high accuracy (typically 0.1\% on macroscopic volume-average quantities). Given the large numbers of simulation results presented hereafter (about 700) and considering the fact that a large number of them are run in finite strain settings up to very large deformations, such discretizations were not achievable for reasonable computation time: as an example, considering 64 random voids with a voxel size of 1/32 of void radius, for an initial porosity of 1\% leads to $1000^3$ voxels in the unit cell. Besides, convergence of the FFT algorithm worsens as voxel size decreases. The effect of discretization has thus been re-evaluated for lower number of voxels for finite strain simulations, for the level of initial porosity used in the following $f = 0.01$. The discretization level is set by the number of voxels along voids radii $\mathcal{D}$. The evolution of stress $\sigma_{11}$ as a function of deformation gradient $F_{11}$ is given in Fig.~\ref{fig2} for different stress triaxialities $T$, hardening moduli $m$ and discretizations $\mathcal{D}$. For $N=1$ void (Fig.~\ref{fig2}a), mesh convergence is already achieved for $\mathcal{D}=3$ except for the very end of the coalescence regime. For a higher number of voids, \textit{e.g.}, $N=16$ voids (Fig.~\ref{fig2}b), convergence of the simulations  was found to be difficult as discretization increases, even though small time steps were used. For large stress triaxiality ($T = 3$), mesh convergence is also achieved for $\mathcal{D}=3$. For lower stress triaxiality ($T = 1$), a stronger effect of discretization is observed for the onset of coalescence (\textcolor{black}{corresponding to the sudden softening behavior}), but increasing discretization leads to convergence issues. Therefore, in the following, all simulation results presented correspond to $\mathcal{D}=3$ for finite strains simulations, keeping in mind that such choice might have a slight effect on coalescence strains for low stress triaxiality, especially as the number of voids increases. For small strains simulations, discretization level is set to $\mathcal{D}=5$. Mesh convergence analysis for particular conditions (not reported here) have shown that such choice leads to an underestimation of the stresses of about 2\%.

\section{Determination of yield surfaces}
\label{ys}
\subsection{Numerical limit analysis}

\begin{figure}[H]
  \centering
  \subfigure[]{\includegraphics[height = 5cm]{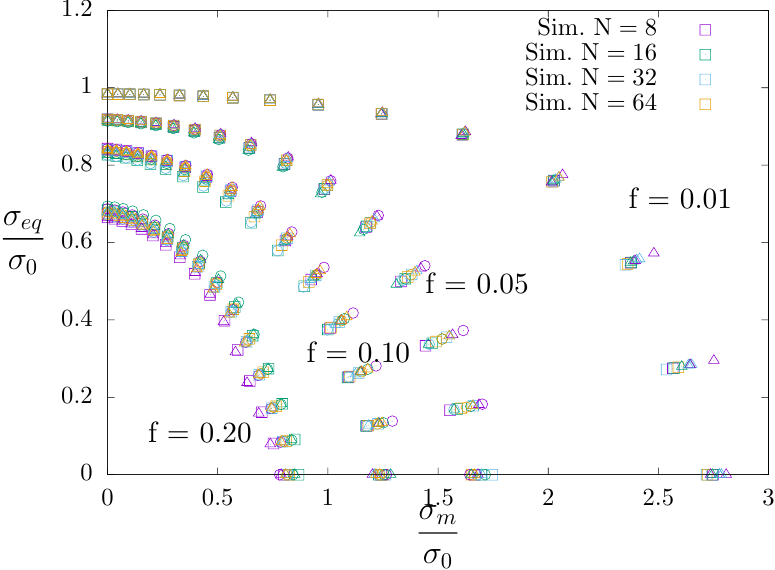}}
  \hspace{0.5cm}
\subfigure[]{\includegraphics[height = 5cm]{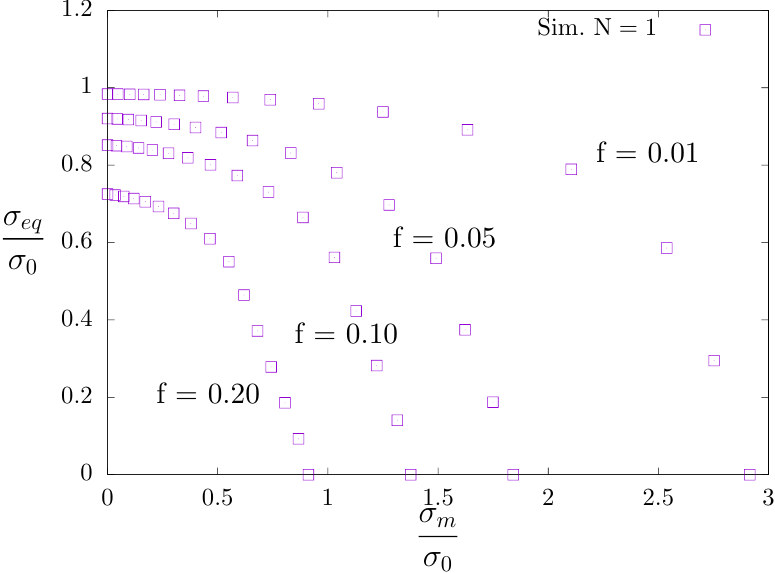}}
\caption{Yield surfaces in  mean stress $\sigma_m$ - von Mises equivalent stress $\sigma_{eq}$ plane determined for different porosities through numerical limit analysis for (a) $N > 1 $ and (b) $N = 1$ voids. In the former case, different point types are used to indicate different void distributions.}
\label{fig3}
\end{figure}

Numerical limit analysis simulations are performed in order to determine the yield surfaces as a function of porosity and void number. Imposed axial strain $\epsilon_{11}$ is increased until convergence of all components of the macroscopic Cauchy stress tensor \cite{bilger2005,fritzen2012}. Values at convergence are used to compute, for a given stress triaxiality, the value of the mean stress $\sigma_m$ and von Mises equivalent stress $\sigma_{eq}$. All results are shown on Fig.~\ref{fig3}, for different void numbers $N > 1$ and three void distributions in Fig.~\ref{fig3}a, and for $N = 1$ in Fig.~\ref{fig3}b as a reference case, defining the yield surfaces. Fig.~\ref{fig3}b shows the typical evolution of the yield surface classically reported in the literature as porosity increases, as well as the appearance of straight parts on the yield surface corresponding to coalescence / inhomogeneous yielding. For higher number of voids in the unit cell (Fig.~\ref{fig3}a), lower stresses are observed for a given porosity and stress triaxiality than for $N = 1$. For low porosity ($f = 0.01$), the effect of void distribution is weak, with similar results for $N = 16$ and $N = 64$, consistently with the fact that void distribution should become irrelevant in the dilute regime. For higher porosities, significant differences are observed as void number is increased, as can be seen by the results obtained for the same void number $N=8$ but different distributions. However, results for $N \geq 32$ are similar, supporting the occurrence of a RVE behavior. 

\begin{figure}[H]
  \centering
  \subfigure[]{\includegraphics[height = 4cm]{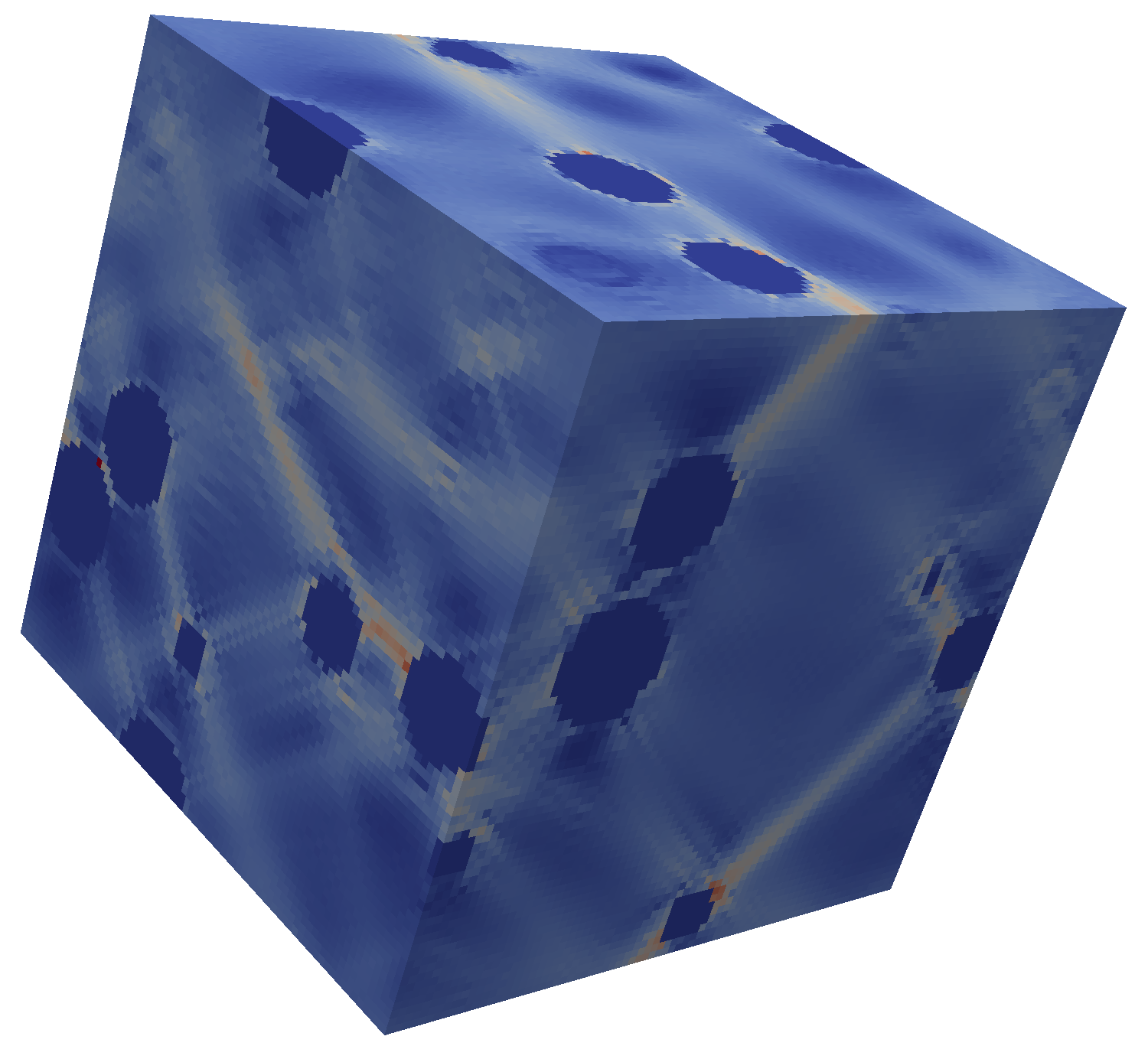}}
    \subfigure[]{\includegraphics[height = 4cm]{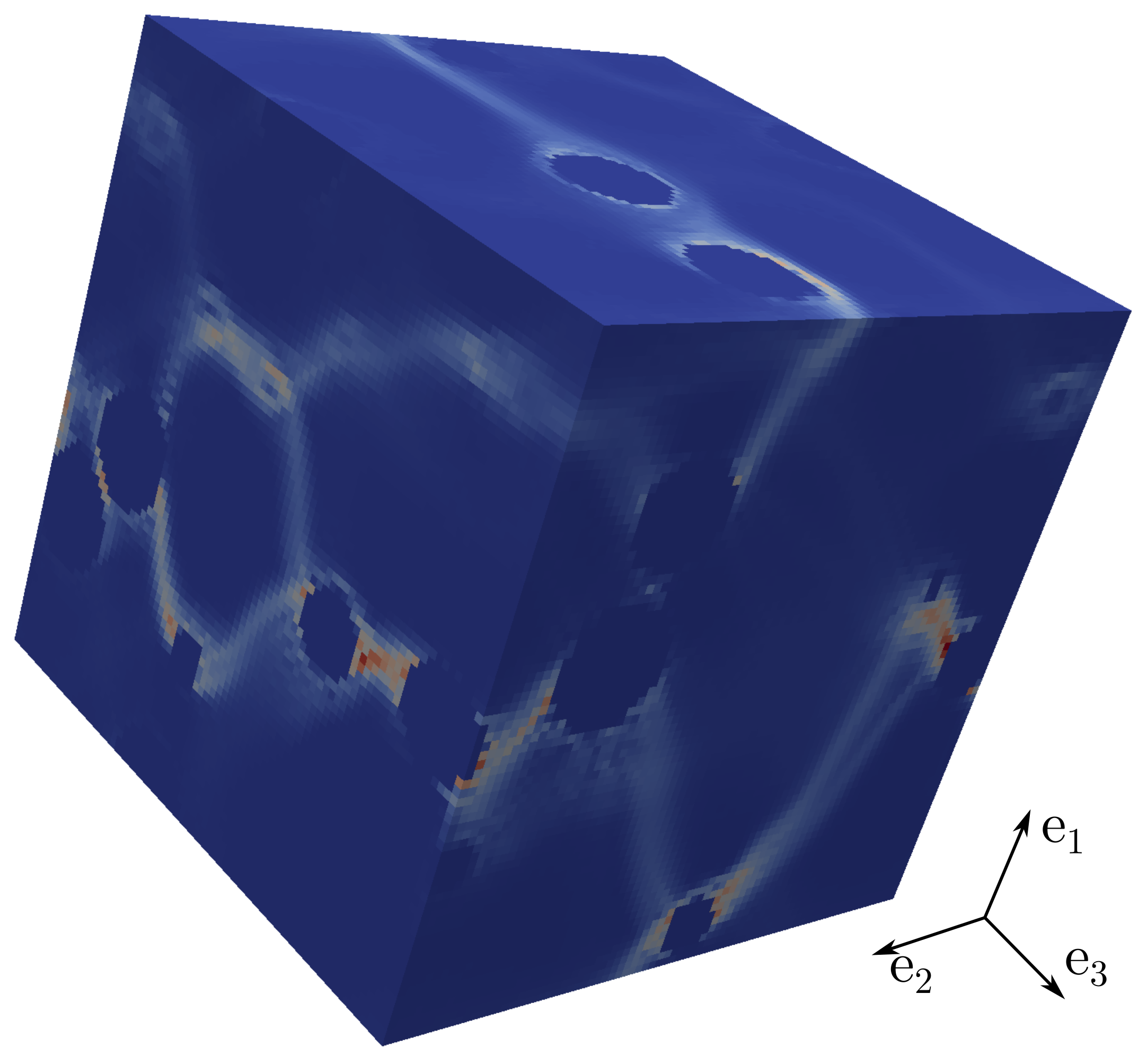}}
\subfigure[]{\includegraphics[height = 4cm]{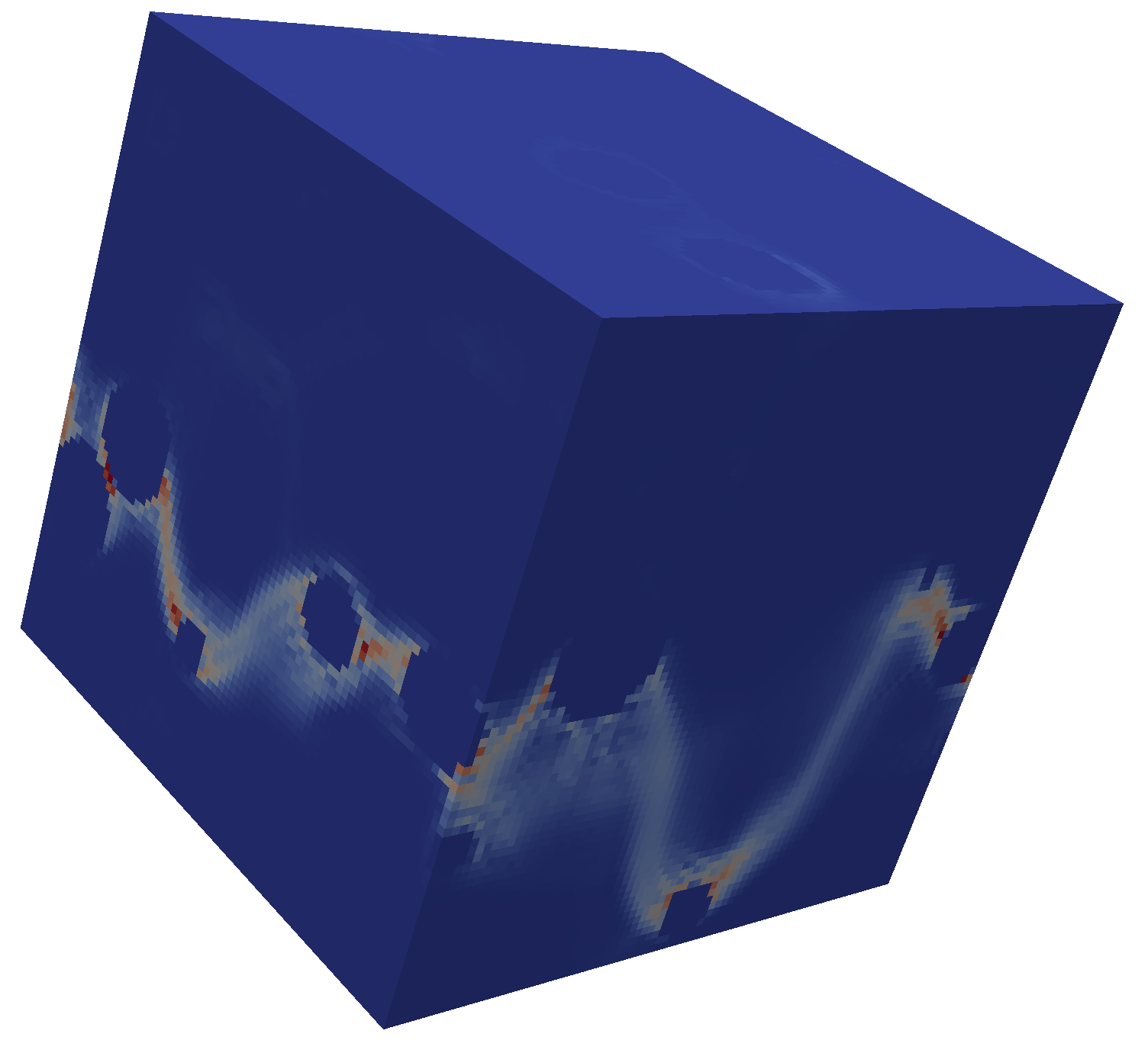}}
\caption{Local accumulated plastic strain $p$ fields for porous unit cells with $N = 32$ voids and porosity $f = 0.1$, for stress triaxiality  (a) $T = 0$, (b) $T = 11/6$ and (c) $T = 13/3$ \textcolor{black}{(Arbitrary units)}}
\label{fig4}
\end{figure}

Fig.~\ref{fig4} shows local accumulated plastic strain fields for a porosity of $f = 0.1$ and $N = 32$ voids, for three stress triaxialities $T = 0 $ (Fig.~\ref{fig4}a), $T = 11/6 $ (Fig.~\ref{fig4}b) and $T = 13/3 $ (Fig.~\ref{fig4}c). At low stress triaxiality, yielding appears rather homogeneous, \textit{i.e.}, plasticity occurs all over the unit cell, even if higher plastic strains can be observed between close voids. For higher stress triaxialities (Fig.~\ref{fig4}b,c), two different behaviors are observed. In both cases, plasticity occurs on a small subset of the unit cell through localization bands linking voids. For $T = 13/3 $, the overall localization layer is perpendicular to the main loading direction, while multiple localization layers are observed for $T = 11/6 $. \textcolor{black}{It should be kept in mind that the localized deformation patterns shown in Fig.~\ref{fig4}b,c may depend on the choice of the unit cell \cite{GLUGE201291} and are constrained by the periodic boundary conditions. A single shear band is thus not possible, hence the appearance of multiple shear bands in Fig.~\ref{fig4}b. A distinction has been done in \cite{tekogluphil} between shear banding and shear coalescence based essentially on the width of the localization band: putting aside that such definition may require some caution, the localized patterns in Fig.~\ref{fig4}b are believed to correspond to local shear coalescence mode.} The results are consistent with 2D results shown in \cite{bilger2005}, and in contrast with what is observed for $N = 1$ voids - a cubic void distribution, where the intermediate regime is not observed.

The results shown in Fig.~\ref{fig3}a  are consistent with the ones from \cite{fritzen2012} for the cases where homogeneous yielding occurs (low stress triaxiality, Fig.~\ref{fig4}a), with a RVE behavior obtained for a number of voids larger than $N = 32$ for the range of porosities considered. The calibrated GTN model proposed in \cite{fritzen2012} is compared to these numerical data in the next section. Moreover, the occurrence of inhomogeneous yielding for random porous isotropic materials with localization layers being either perpendicular or slanted with respect to the main loading directions is consistent with the hypothesis of the model developed in \cite{KERALAVARMA2017100,REDDI2019190}, which is thus also used in the following.

\subsection{Comparisons to theoretical models}

\begin{figure}[H]
  \centering
  \subfigure[]{\includegraphics[height = 5cm]{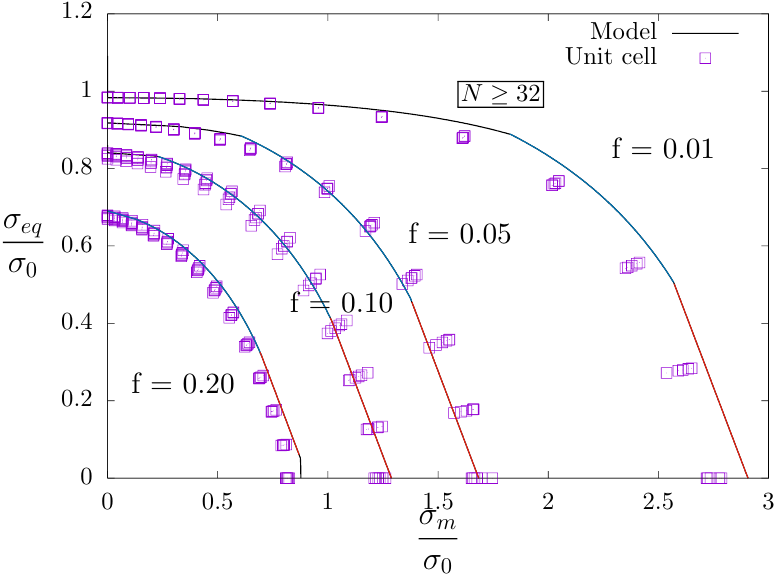}}
  \hspace{0.5cm}
  \subfigure[]{\includegraphics[height = 5cm]{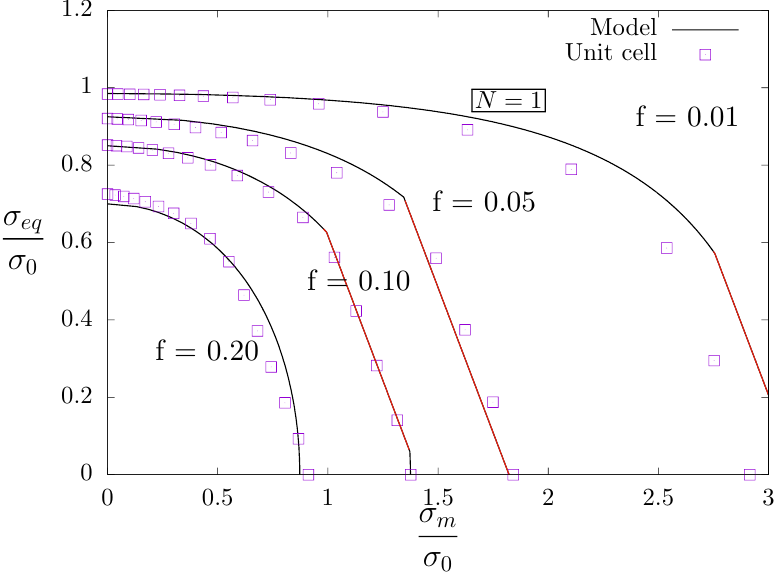}}\\
    \subfigure[]{\includegraphics[height = 5cm]{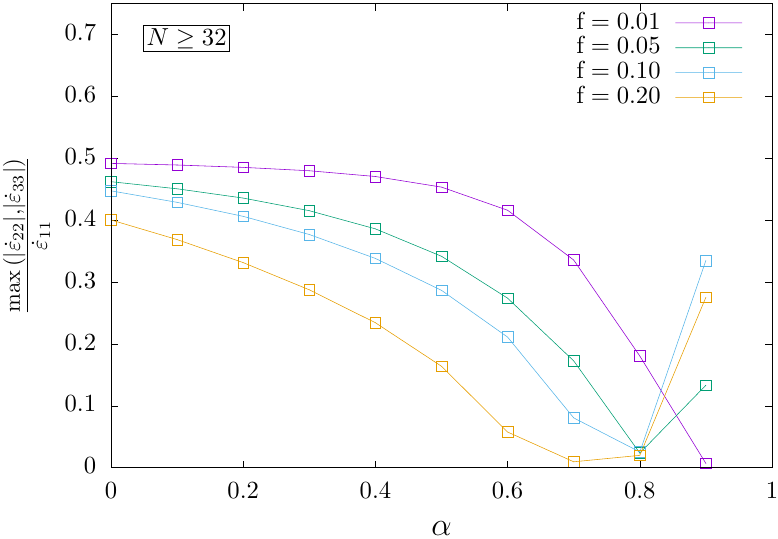}}
  \hspace{0.5cm}
\subfigure[]{\includegraphics[height = 5cm]{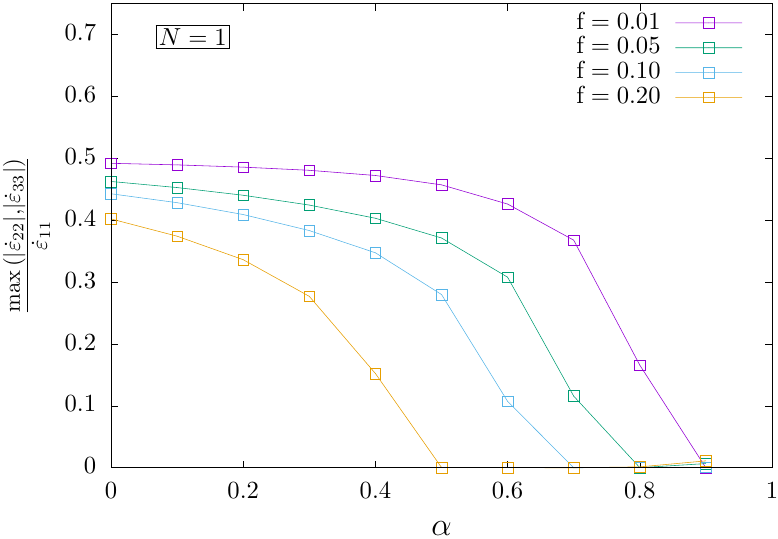}}
\caption{(a,b) Yield surfaces in the plane mean stress $\sigma_m$ - von Mises equivalent stress $\sigma_{eq}$ \textcolor{black}{and (c,d) associated strain rates as a function of $\alpha = \sigma_{22} / \sigma_{11} = \sigma_{33} / \sigma_{11}$} determined for different porosities through numerical limit analysis for (a,c) $N  \geq 32 $ and (b,d) $N = 1$ void. (a,b) Comparisons to model predictions:  the lines correspond to the model with (a) $\{q_1,q_2,\gamma\} = \{1.69 - f,0.92,1.25\}$ and (b) $\{q_1,q_2,\gamma\} = \{1.5,0.92,1.\}$. Black, blue and red lines correspond to the activation of Eqs.~\ref{f1},~\ref{f2},~\ref{f3}, respectively.}
\label{fig5}
\end{figure}

Two yield criteria proposed in the literature are compared to the numerical results presented in the previous section. The first one is the calibrated GTN model proposed in \cite{fritzen2012} for homogeneous yielding of random porous isotropic materials, which is:
\begin{equation}
  \mathcal{F}_1 = \left(\frac{\sigma_{eq}}{\sigma_0}\right)^2 + 2\,q_1\,f\,\cosh{\left(\frac{3}{2}q_2\frac{\sigma_m}{\sigma_0}\right)} - 1 - (q_1\,f)^2 \leq 0
  \label{f1}
\end{equation}
where the parameters $q_1$ and $q_2$ have been calibrated to:
\begin{equation}
  \begin{aligned}
    q_1 &= 1.69 - f \\
    q_2 &= 0.92
  \end{aligned}
  \label{q1q2}
\end{equation}
The second yield criterion is the coalescence criterion detailed in \cite{KERALAVARMA2017100,REDDI2019190} assuming an isotropic distribution of spherical voids in a perfectly plastic isotropic material. This criterion is composed of two yield surfaces. The first one correspond to shear assisted coalescence:
\begin{equation}
    \mathcal{F}_2 = 3\left[\left(\frac{\sigma_1 - \sigma_2}{2\sigma_0}\right)^2 - \eta^2 \left(\frac{\sigma_1 + \sigma_2}{2\sigma_0}\right)^2\right] + 2\,f_b\,\cosh{\left(\kappa\frac{\sigma_1+\sigma_2}{2\sigma_0}\right)} - 1 - f_b^2 \leq 0
\label{f2}
\end{equation}
where $\sigma_1$ and $\sigma_2$ are two different \textcolor{black}{unequal} principal stresses, $f_b$ the porosity in the coalescence layer, and $\{\eta,\kappa\}$ parameters defined as:
\begin{equation}
  \begin{aligned}
    \kappa &= \beta(1 + \eta)  &\eta &= \min{\left(   \frac{f_b \beta^2}{3 - f_b \beta^2}, \textcolor{black}{\left|\frac{\sigma_1 - \sigma_2}{\sigma_1 + \sigma_2
      } \right|  } \right) }\\
    \beta &= \sqrt{\frac{5}{6}}\log{\left(\frac{1}{f_b}\right)}\left(   \sqrt{b^2 + 1} - \sqrt{b^2 + f_b^2}  + b \log{\left( \frac{b + \sqrt{b^2 + f_b^2} }{f_b (b + \sqrt{b^2 + 1})}      \right)}      \right)^{-1}\\
    b &=  \sqrt{\frac{1}{3} + \frac{5 A}{24 f_b}} & A &= \frac{1 + f_b - 5f_b^2 + 3 f_b^3}{12}
    \end{aligned}
  \end{equation}
The second yield surface corresponds to coalescence by internal necking:
\begin{equation}
  \mathcal{F}_3   =  2 f_b \cosh{\left(\beta \frac{\sigma_1}{\sigma_0}\right)}  -1 - f_b^2  \leq 0
  \label{f3}
\end{equation}
and is a refinement of the Thomason coalescence criterion \cite{thomason85a}. Eqs.~\ref{f2} and \ref{f3} require some comments. Both of them correspond in fact to three yield surfaces due to the possible permutations on the principal stresses $\sigma_i$. As normality is assumed for plastic flow, Eq.~\ref{f3} leads to uniaxial straining conditions in the direction sets by the maximal principal stresses, consistently with the internal necking deformation mechanism. On the contrary, as shown in \cite{KERALAVARMA2017100,REDDI2019190}, associated plastic flow of Eq.~\ref{f2} is not uniaxial as a result of the homogenization of a localization porous layer under combined tension and shear at the microscale. In Eqs.~\ref{f2} and \ref{f3}, the key parameter is the porosity in the coalescence layer $f_b$. \textcolor{black}{This parameter has been introduced in \cite{torki} based on the following argument. Most of the recent theoretical models for coalescence / inhomogeneous yielding are derived considering a cylindrical unit cell (of radius $L$) containing a cylindrical void (of radius $R$). The parameter $\chi = R/L$ appears naturally in the derivation as plastic flow is considered to be localized in the intervoid ligament, and is equal to $\chi = \sqrt{f_b}$ where $f_b$ is the porosity in the coalescence layer (of height $2R$). As extending coalescence criteria considering more complex geometries has been found to be challenging \cite{barriozJAM}, formulas derived using cylindrical unit cell are still used replacing $\chi$ by $f_b^2$.}, which can be related through scaling analysis to the global porosity $f$ through the relation $f_b \sim f^{2/3}$. The prefactor of this relation might depend on void distribution, thus in the following:
\begin{equation}
  f_b = \left(\gamma f  \right)^{2/3}
  \label{fb}
  \end{equation}
Eqs.~\ref{f1},~\ref{f2},~\ref{f3} can be used simultaneously to build a multi-surface yield criterion, as done in \cite{KERALAVARMA2017100,REDDI2019190} using $\{q_1,q_2,\gamma \} = \{1,1,1\}$, to model isotropic porous materials containing random distributions of voids. However, this model has not yet been compared to numerical data based on random distributions of voids, as the comparisons were restricted in \cite{KERALAVARMA2017100,REDDI2019190} to cubic array of voids. This model is thus compared to the results obtained in this study for $N \geq 32$ voids in Fig.~\ref{fig5}a, taking the GTN parameters to the ones calibrated in \cite{fritzen2012} (Eq.~\ref{q1q2}), and calibrating the parameter $\gamma = 1.25$. A very good agreement is obtained in the range of porosity $f \in [0.01:0.20]$, with a slight overestimation for the lowest porosity. For all porosities, different yield surfaces (Eqs.~\ref{f1},~\ref{f2},~\ref{f3}) are activated depending on the stress triaxiality ratio, from homogeneous yield at small values of $T$ to coalescence through internal necking at high $T$, where shear-assisted coalescence is activated in between. This is consistent with the observations of plastic strain field reported in Fig.~\ref{fig4} where three different deformation modes were observed. For very high porosity, \textit{i.e.}, $f = 0.20$, no homogeneous yield is predicted by the model even for $T = 0$.

Fig.~\ref{fig5}b shows the same comparison for the case of a single void in the unit cell, thus corresponding to the homogenization of a cubic array of spherical voids where the principal axis of the void distribution are aligned with the axis of the loading. A good agreement is observed by using the parameters $\{q_1,q_2,\gamma\} = \{1.5,0.92,1.\}$ up to a porosity of 10\% and not accounting for Eq.~\ref{f2}, as for this particular distribution, only homogeneous yielding and coalescence by internal necking are observed, as the void distribution dictates the plane on which coalescence can occur.

\textcolor{black}{Figs.~\ref{fig5}c,d show the corresponding transverse strain rates (normalized by the imposed strain rate $\dot{\varepsilon}_{11}$) for the different porosities, as a function of the parameter $\alpha = \sigma_{22} / \sigma_{11} = \sigma_{33} / \sigma_{11}$. For $N=1$ (Fig.~\ref{fig5}d), a clear transition is observed for large porosity values / large values of $\alpha$ (which corresponds to large mean stress $\sigma_m$ by Eq.~\ref{eqT}) towards uniaxial straining. This is consistent with the occurrence of coalescence by internal necking (Fig.~\ref{fig5}b). For $N \geq 32$ (Fig.~\ref{fig5}c), uniaxial straining conditions do appear, but in a narrower range than for $N=1$, which is also consistent with the model (Fig.~\ref{fig5}a) restricting coalescence by internal necking (red lines) to limited parts of the yield surfaces. Interestingly, for very high stress triaxialities ($\alpha = 0.9$), uniaxial straining conditions stop to apply, which is not predicted by the model.} 

 The overall good agreement between the model proposed in \cite{KERALAVARMA2017100,REDDI2019190} and the numerical results regarding both macroscopic yield surface and deformation mechanisms once properly calibrated allows to use it to perform evolution problems, which is assessed in the next section.

\section{Finite strain evolutions}
\label{fs}
\subsection{Unit cell results}

\begin{figure}[H]
  \centering
  \subfigure[]{\includegraphics[height = 5cm]{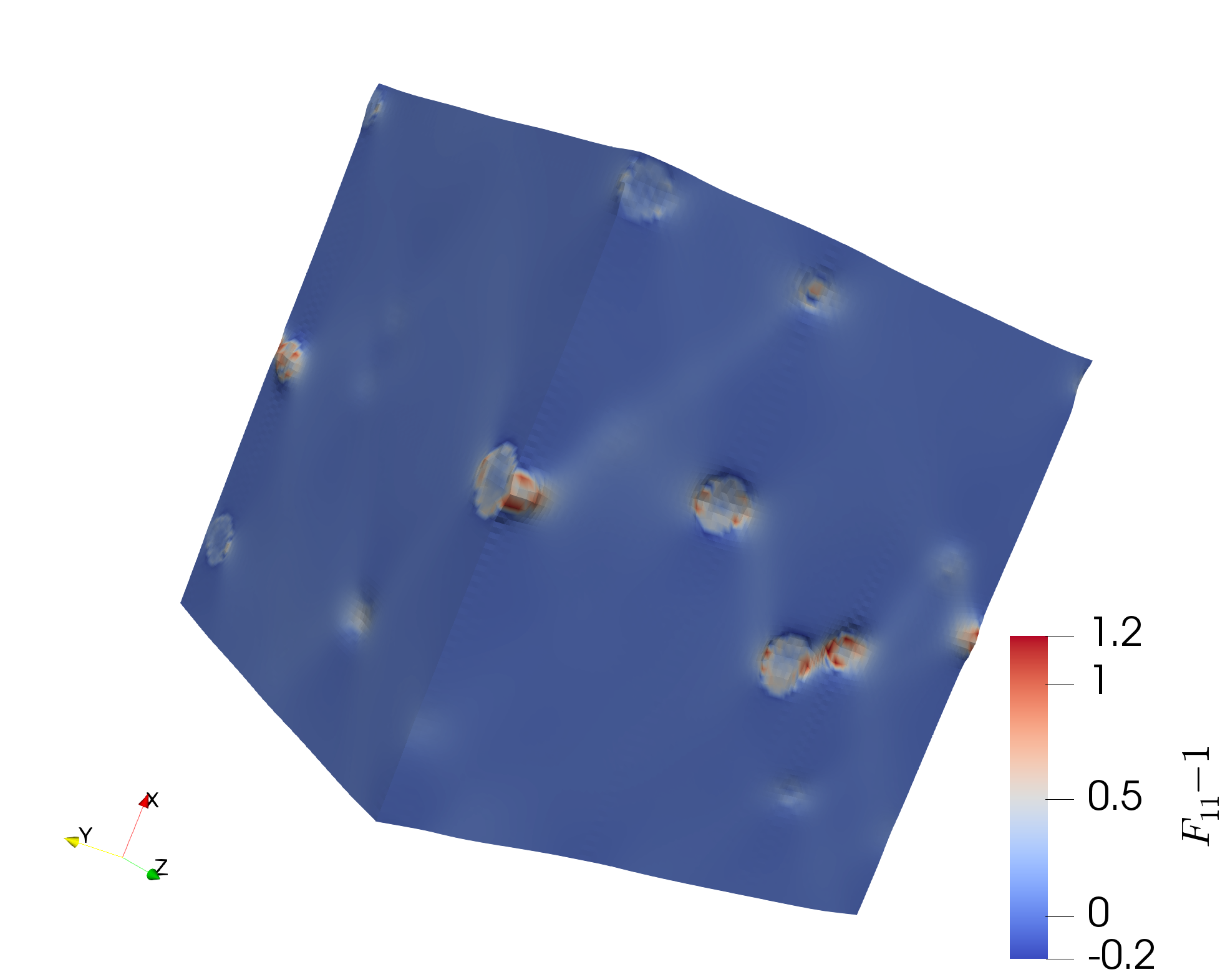}}
  \hspace{0.5cm}
\subfigure[]{\includegraphics[height = 5cm]{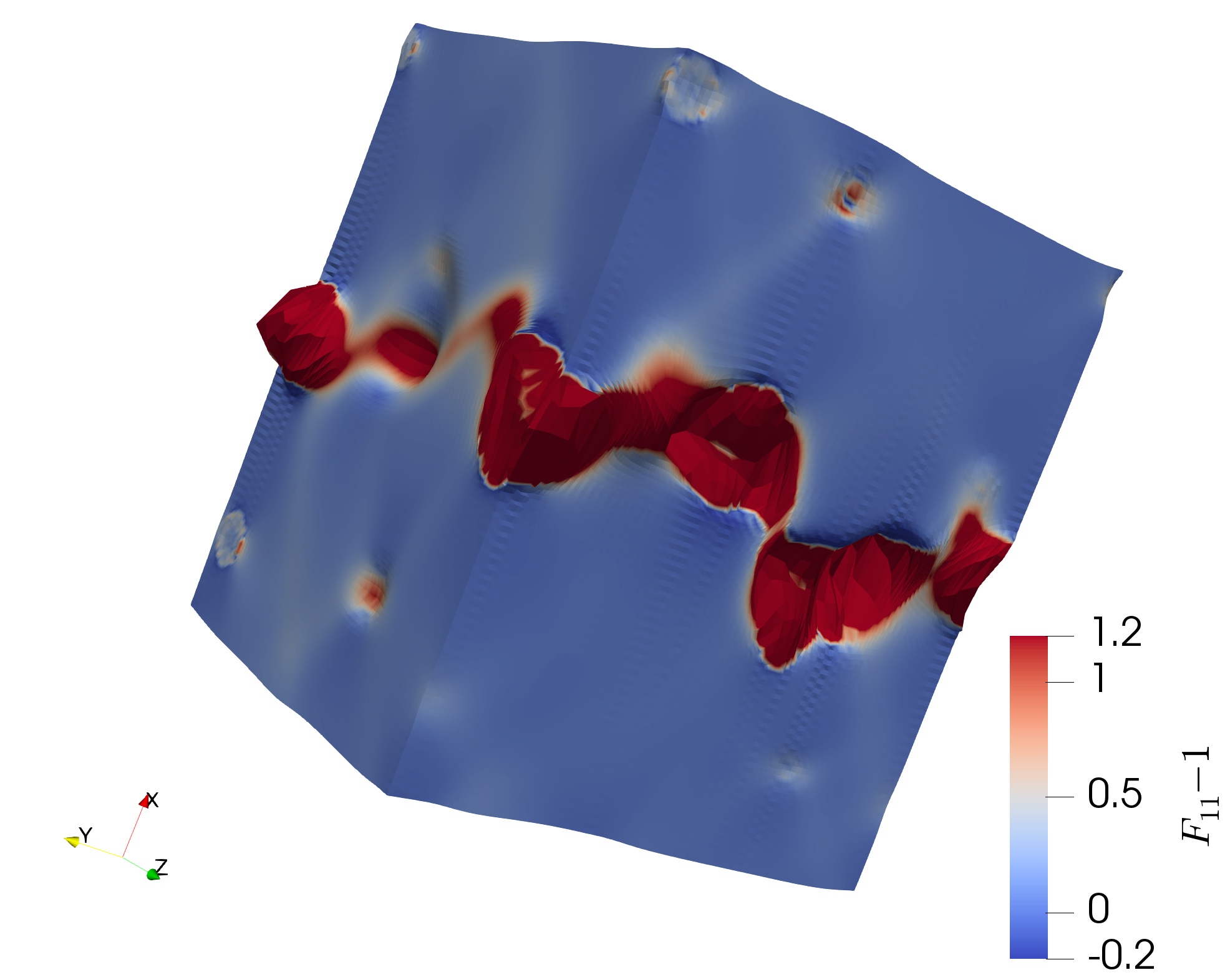}}
\caption{Examples of deformed unit cells containing 32 voids, for a stress triaxiality of $T = 2$, in the (a) homogeneous and (b) inhomogeneous deformation regimes}
\label{fig7}
\end{figure}

Finite strain porous unit cells results are described in this section for the loading conditions and discretization described in Section~\ref{num}. For all simulations, the initial porosity is set to $f = 0.01$. Different hardening moduli $m \in [0.0;0.1;0.2]$, void numbers $N \in [1:64]$ and stress triaxialities $T \in [1:3]$ are considered. Examples of deformed unit cells containing 32 voids are shown in Fig.~\ref{fig7}, for a stress triaxiality of $T = 2$, in the homogeneous and inhomogeneous deformation regimes. 

\begin{figure}[H]
  \centering
  \subfigure[]{\includegraphics[height = 5cm]{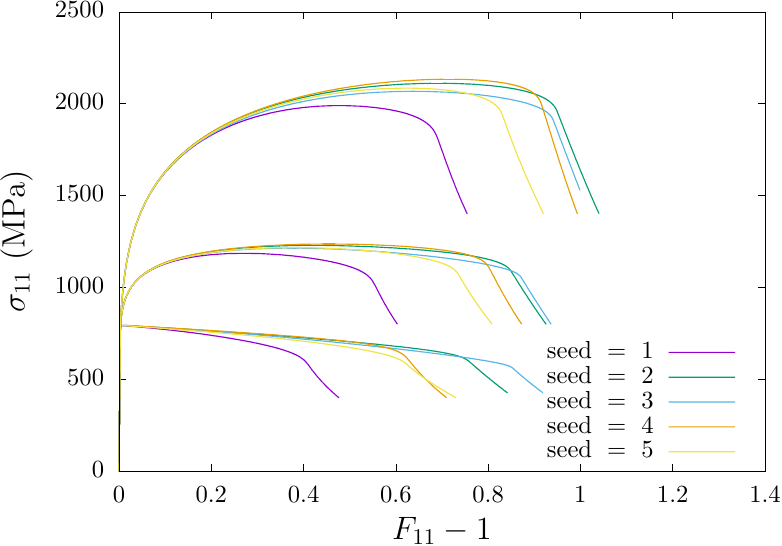}}
  \hspace{0.5cm}
  \subfigure[]{\includegraphics[height = 5cm]{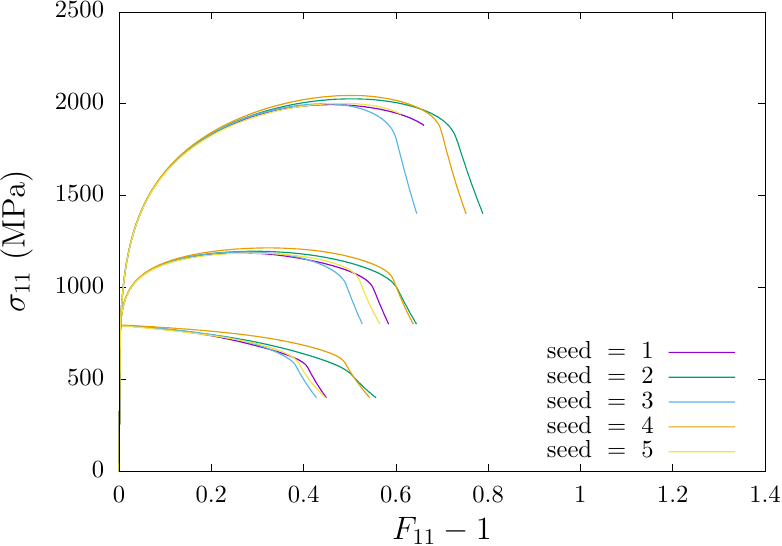}}
    \subfigure[]{\includegraphics[height = 5cm]{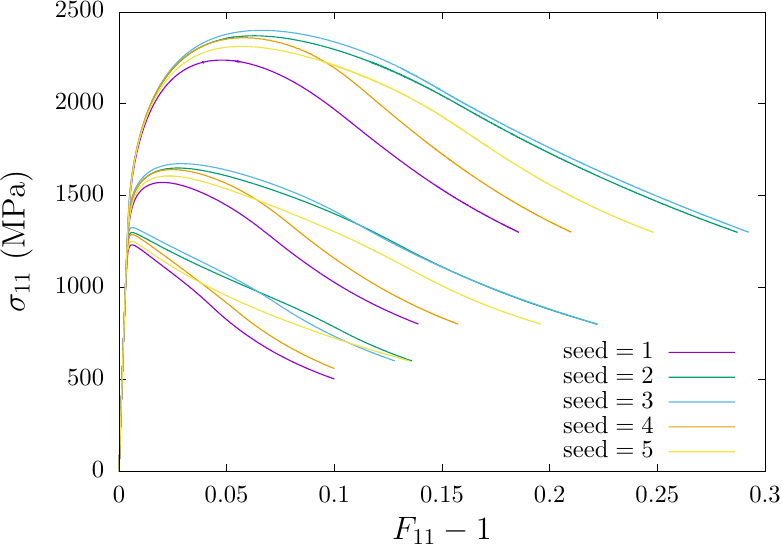}}
  \hspace{0.5cm}
\subfigure[]{\includegraphics[height = 5cm]{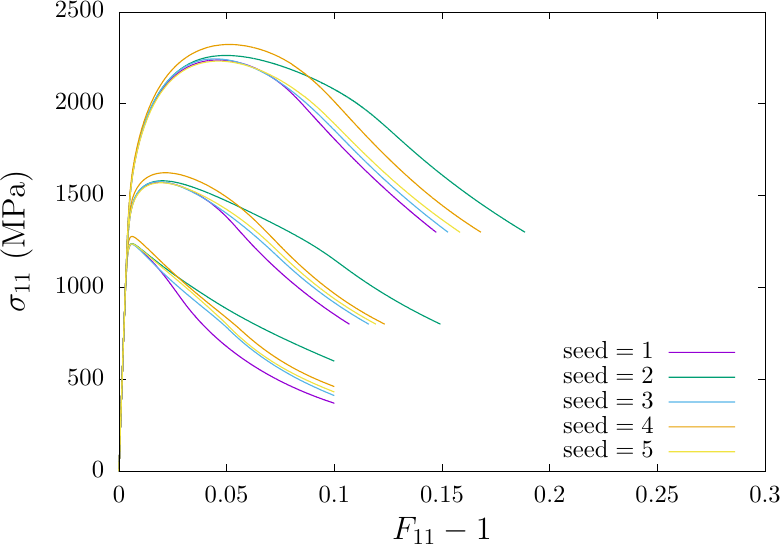}}
\caption{Evolution of the macroscopic Cauchy stress $\sigma_{11}$ as a function of the imposed macroscopic deformation gradient $F_{11}$ for porous unit cells with different hardening moduli $m \in [0.0;0.1;0.2]$ and distribution of voids, for stress triaxiality ratio (a,b) $T = 1$ and (c,d) $T = 3$, for (a,c) $N = 4$ and (b,d) $N = 16$ voids}
\label{fig6}
\end{figure}

Fig.~\ref{fig6} shows the evolution of stress $\sigma_{11}$ as a function of the deformation gradient $F_{11}$ for different stress triaxialities $T$, hardening moduli $m$ and distributions of voids (through different seeds of the random generator). For $T=1$ (Fig.~\ref{fig6}a,b), the mechanical response clearly depends on void distributions, for a given porosity $f=0.01$ and few voids $N=4$ (Fig.~\ref{fig6}a). This is particularly true for the coalescence regime, which corresponds to the increase of softening, for low hardening matrix material, although a slight effect can also be observed in the growth regime. For $N=16$ voids (Fig.~\ref{fig6}b), the dependence of the mechanical response to the void distribution decreases. As could have been expected, the value of the slope in the coalescence regime depends on the number of voids in the unit cell: as coalescence corresponds to the localization of deformation in a layer of height approximately equal to one void diameter (as can be seen on Fig.~\ref{fig7}b), the higher the void number, the lower the void diameter for a given porosity, and the thinner the coalescence layer that account for all the deformation of the unit cell. Thus the macroscopic deformation rate $\dot{E}$ in a unit cell of size $L$ scales as $\dot{E} \sim (R/L) \dot{\epsilon}$, where  $\dot{\epsilon}$ is the strain rate in the coalescence layer. For $R \rightarrow 0$, the coalescence slope would be infinite. All these observations made for the case of moderate stress triaxiality $T = 1$ are similar for higher stress triaxiality $T=3$, as shown on Fig.~\ref{fig6}c,d.

\begin{figure}[H]
  \centering
  \subfigure[]{\includegraphics[height = 5cm]{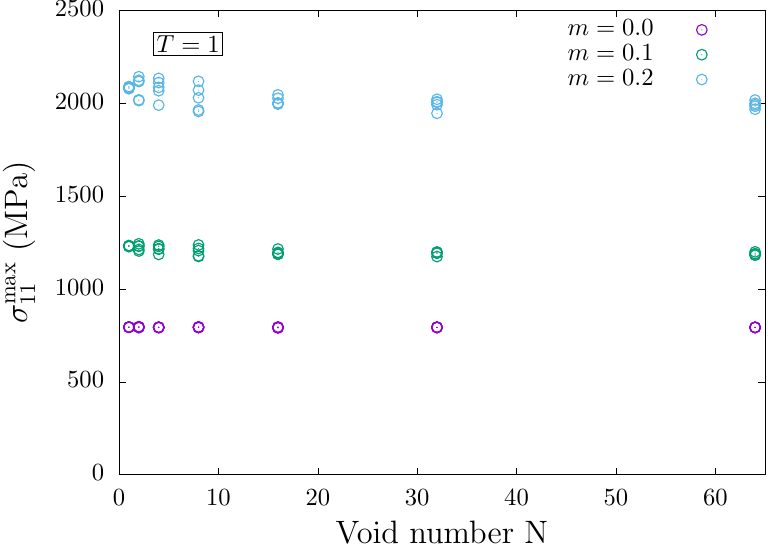}}
  \hspace{0.5cm}
  \subfigure[]{\includegraphics[height = 5cm]{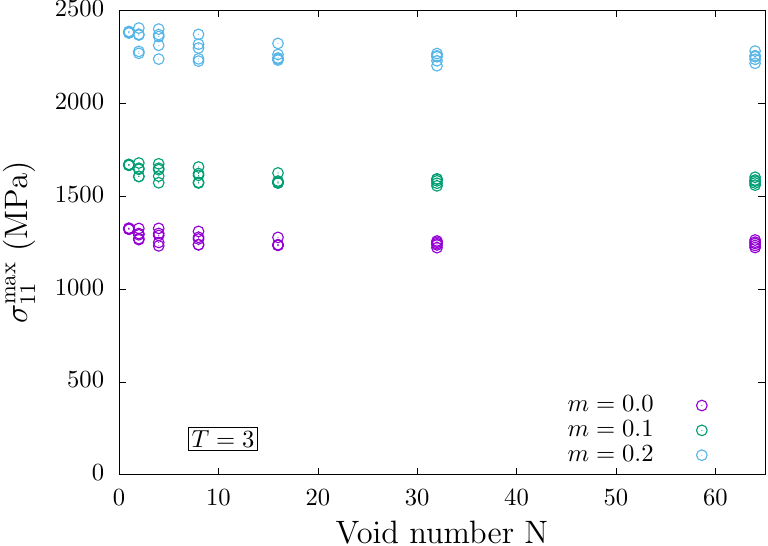}}
    \subfigure[]{\includegraphics[height = 5cm]{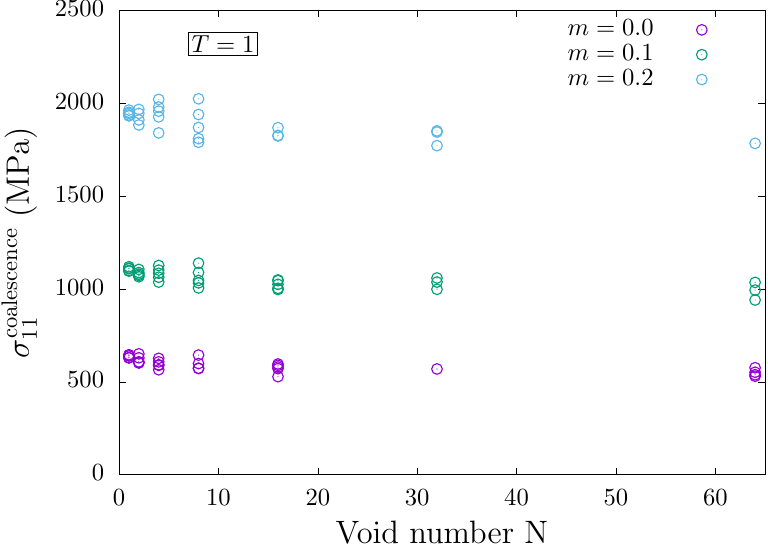}}
  \hspace{0.5cm}
\subfigure[]{\includegraphics[height = 5cm]{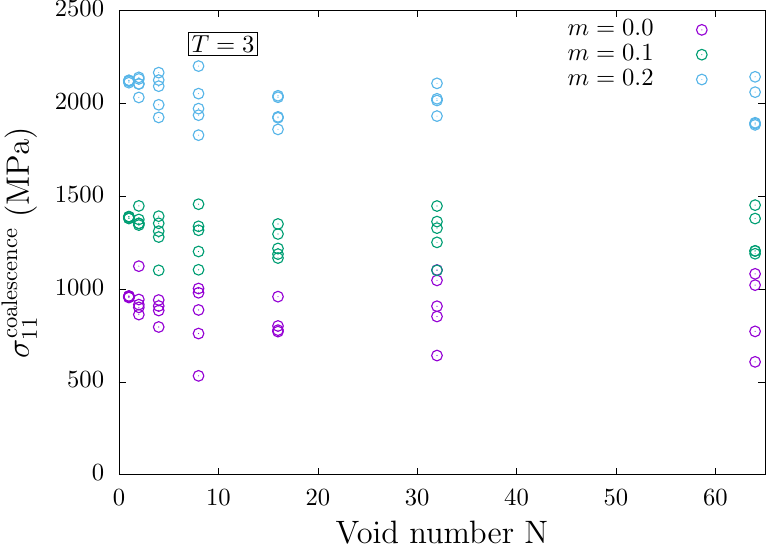}}
\caption{Evolution of maximal stress $\sigma_{11}^{\mathrm{max}}$ (a,b) and coalescence stress $\sigma_{11}^{\mathrm{coalescence}}$ (c,d) as a function of void number, for stress triaxiality  $T=1$ (a,c) and $T=3$ (b,d), for different hardening moduli $m$ }
\label{fig8}
\end{figure}

Fig.~\ref{fig6} shows that, as void number increases, a less ductile behavior is observed, with an earlier softening and onset of coalescence. For a number of voids higher than $N=16$, the macroscopic behavior tends to saturate for the maximal and coalescence stresses, indicating the occurrence of a Representative Volume Element (RVE). In order to assess in more details the convergence towards a RVE as a function of voids number, maximal stress $\sigma_{11}^{\mathrm{max}}$ and coalescence stress $\sigma_{11}^{\mathrm{coalescence}}$ are shown on Fig.~\ref{fig8} for different void numbers, hardening moduli and stress triaxialities. \textcolor{black}{As discussed in \cite{koplik}, two different phases can be distinguished in porous unit cell simulations at constant stress triaxiality under axisymetric loading conditions.  The first one corresponds to macroscopic triaxial straining. Then a transition is observed towards macroscopic uniaxial straining as a result of inhomogeneous yielding. Therefore, the onset of} coalescence stress is defined as the value of the stress when coalescence \textcolor{black}{starts}: $\max{\left(\dot{F}_{22},\dot{F}_{33} \right) \leq 0.05 \dot{F}_{11}}$. Regarding the maximal stress as a function of the void number (Fig.~\ref{fig8}a,b), a RVE is obtained for $N \geq 32$ for the different hardening exponents and stress triaxialities. For the coalescence stress (Fig.~\ref{fig8}c,d), it should be noticed that the mean coalescence stress is found to be only slightly dependent on the void number. For moderate stress triaxiality $T=1$, a RVE is also obtained for $N \geq 32$. For higher stress triaxiality $T=3$, more variability is observed of the coalescence stress as a function of void distribution, even for large number of voids ($N = 64$), which prevent to conclude about the attainment of a RVE. Such variability can be related to mesh convergence, as described in Fig.~\ref{fig2}. Another cause might be that, for cubic unit cell and high stress triaxiality conditions, coalescence can occur along different planes whereas the coalescence plane is perpendicular to the main loading direction for lower stress triaxiality.

The results obtained in this section about the finite strain computational homogenization of isotropic materials containing (pseudo-)random distributions of initially spherical voids lead to the definition of a RVE behavior for void number higher than about 32, up to the onset of coalescence. Such result is in accordance with the small strain computational homogenization performed in \cite{fritzen2012}, extending their result for finite strain. All the unit cells results obtained for $N \geq 32$ are plotted on Fig.~\ref{fig9} for the different stress triaxialities, hardening moduli and voids distributions. A RVE behavior is rather clearly observed for $T=2$ and $T=3$. For $T=1$, the onset of coalescence still depend slightly on void distribution. The dependence is much weaker than for lower void number, as shown in Fig.~\ref{fig8}, and may come either from discretization effect or either from a higher number of voids needed to get a RVE for low stress triaxiality. However, as discussed in Section.~\ref{num}, it was not possible to go for higher discretization / higher number of voids in this study due to severe convergence issues. Based on the results shown in Fig.~\ref{fig9}, the next step is to propose a homogenized model allowing to reproduce these unit cells results, which is done the next section. 

\subsection{Homogenized model}

\begin{figure}[H]
  \centering
  \subfigure[]{\includegraphics[height = 5cm]{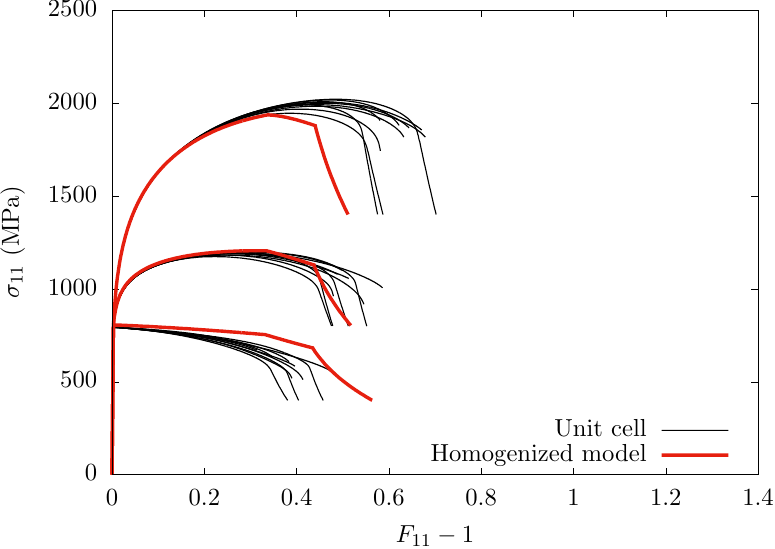}}
  \hspace{0.5cm}
  \subfigure[]{\includegraphics[height = 5cm]{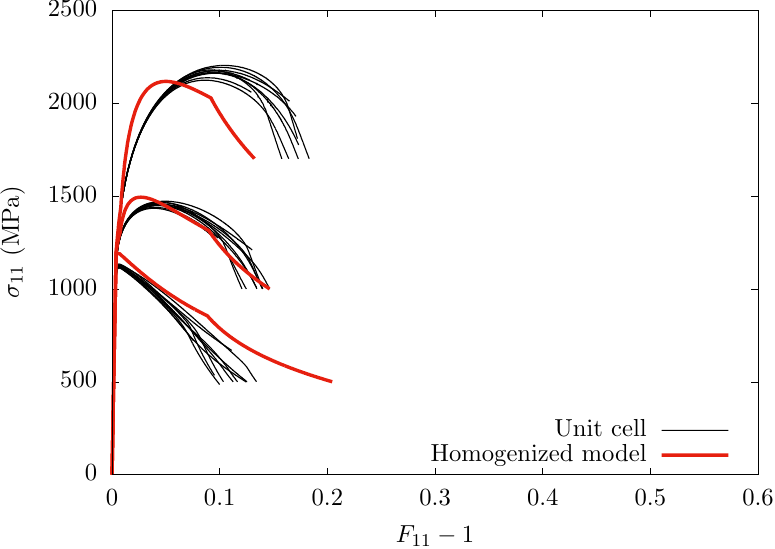}}\\
    \subfigure[]{\includegraphics[height = 5cm]{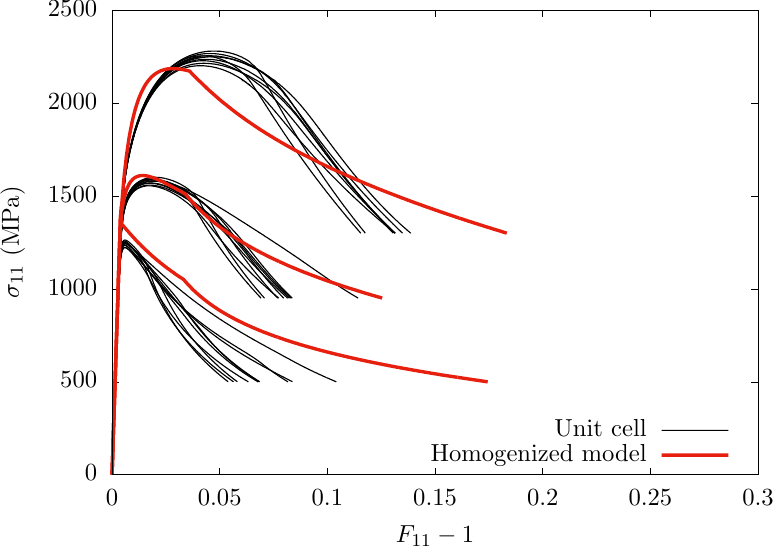}}
\caption{Evolution of the macroscopic Cauchy stress $\sigma_{11}$ as a function of the imposed macroscopic deformation gradient $F_{11}$ for porous unit cells with different hardening moduli $m \in [0.0;0.1;0.2]$ and distribution of voids, for stress triaxiality ratio (a) $T = 1$, (b) $T = 2$, and (c) $T = 3$. Comparison between unit cells results for $N \geq 32$ to homogenized model predictions}
\label{fig9}
\end{figure}

A homogenized model is proposed for random porous isotropic materials in order to reproduce the results shown in Fig.~\ref{fig9}. The three yield surfaces (Eqs.~\ref{f1},~\ref{f2},~\ref{f3}) described in Section~\ref{ys} and shown to be in good agreement with numerical limit analysis results are used simultaneously:
\begin{equation}
  \mathcal{F}\left(\underline{\sigma}, \sigma_0, f    \right) = \max{ \left( \mathcal{F}_1, \mathcal{F}_2, \mathcal{F}_3      \right)}  = 0
  \label{sstar}
\end{equation}
which defines implicitly an equivalent stress $\sigma_{\star} = \sigma_0$ that is used to define a yield criterion, following \cite{BCCF}:
\begin{equation}
  \phi = \sigma_{\star} - R(p)
  \label{ce1}
\end{equation}
where $R(p)$ corresponds to the isotropic hardening of the matrix material, and $p$ an average plastic strain. Plastic flow is assumed to obey normality:
\begin{equation}
  \textcolor{black}{\dot{\underline{\varepsilon}}_p = \dot{\Lambda} \frac{\partial \sigma_{\star}}{\partial \underline{\sigma}} =  \dot{\Lambda} \underline{n}}
 \label{ce2}
\end{equation}
The direction of the plastic flow $\underline{n}$ is computed using Euler theorem for homogeneous function, leading to:
\begin{equation}
    \underline{n} = - \left(\frac{\partial \mathcal{F}}{\partial \sigma_{\star}} \right)^{-1} \frac{\partial \mathcal{F}}{\partial \underline{\sigma}}
  \end{equation}
where $\mathcal{F}$ is given by Eq.~\ref{sstar}. The expression for GTN model and the inhomogeneous yield criterion can be found elsewhere \cite{BCCF,REDDI2019190}. \textcolor{black}{The modelling of hardening of the matrix material is taken from \cite{VISHWAKARMA2019135}}:
\begin{equation}
  \dot{\Lambda} = \left\{ \begin{array}{ll}
	(1 - f) \dot{p} & \ \mathrm{for\ homogeneous\ yielding}\\
        (kf^{1/3} - f) \dot{p} & \mathrm{for\ inhomogeneous\ yielding}
  \end{array}\right.
  \label{eqhard}
\end{equation}
\textcolor{black}{that gives the evolution of the average plastic strain in the matrix $p$, where $k$ is a parameter to calibrate.} The evolution of the void volume fraction is based on plastic incompressibility of the matrix material \cite{benzergaleblond}:
\begin{equation}
  \dot{f} = (1 - f) \, \mathrm{trace}\left( \dot{\underline{\varepsilon}}_p \right)
   \label{ce3}
\end{equation}
For the sake of simplicity, voids are assumed to remain spherical. As discussed previously, the inhomogeneous yield criterion strongly depends on the parameter $f_b$ which is the porosity in the coalescence layer. The prefactor $\gamma$ in Eq.~\ref{fb} was calibrated for cubic unit cells containing random distributions of spherical voids, while in finite strain the unit cells become tetragonal and voids either prolate (\textit{e.g.,} for $T = 1$) or oblate (\textit{e.g.,} for $T = 3$). The effect of void shape will be discussed in the next section. In order to account for the effect of unit cell shape on coalescence by internal necking, the following argument is used. Considering a tetragonal unit cell of size $L$ and aspect ratio $\lambda$ containing a single spherical void of radius $R$, the porosity is equal to $f = (4 \pi R^3)/(3 \lambda L^3)$. Assuming coalescence by internal necking, the porosity in the coalescence layer is $f_b = (4 \pi R^2)/(3 L^2)$. The relation between these two porosities is thus:
\begin{equation}
  f_b = (\gamma \lambda f)^{2/3}
  \label{fb2}
\end{equation}
where $\gamma$ is the prefactor determined for a cubic unit cell $\lambda = 1$. Eq.~\ref{fb2} is thus used in Eq.~\ref{f3} to account for the evolution of the unit cell shape. Concerning the yield surfaces described by Eq.~\ref{f2}, it is unclear how Eq.~\ref{fb} should be corrected, thus no correction is made.\\

A classical implicit integration scheme is used to integrate the constitutive equations defined by Eqs.~\ref{ce1},~\ref{ce2},~\ref{ce3}. The numerical implementation has been done in the \texttt{MFront} code generator \cite{mfront}, and follows the implementation detailed in \cite{SCHERER2019135}. Fig.~\ref{fig9} shows, for three different stress triaxiality ratio $T \in [1;2;3]$ and hardening moduli $m \in [0.0;0.1;0.2]$ the unit cells results for $N \geq 32$ as well as the results from the homogenized model. A good agreement is observed, without any fitting parameter other than the ones calibrated on yield surfaces and the corrections of Eq.~\ref{fb2} for low stress triaxiality / low hardening modulus \textcolor{black}{where the parameter $k$ from Eq.~\ref{eqhard} has little influence. For high hardening modulus $m = 0.2$ / high stress triaxiality, a good agreement is obtained by calibrating the parameter $k=2$. This value is higher than the one proposed in \cite{VISHWAKARMA2019135} where $k=1$ was used. Incorporating hardening in a quantitative manner for porous constitutive equations has always been a challenge, that has been overcome either through heuristic corrections of the yield criteria \cite{pardoen}, or by considering physically based modeling \cite{MORIN2017167}. Although using Eq.~\ref{eqhard} leads to satisfactory predictions, more studies are required on the hardening of random porous materials under inhomogeneous yielding.}

\section{Discussion and Perspectives}
\label{disc}

\begin{figure}[H]
  \centering
  \subfigure[]{\includegraphics[height = 5cm]{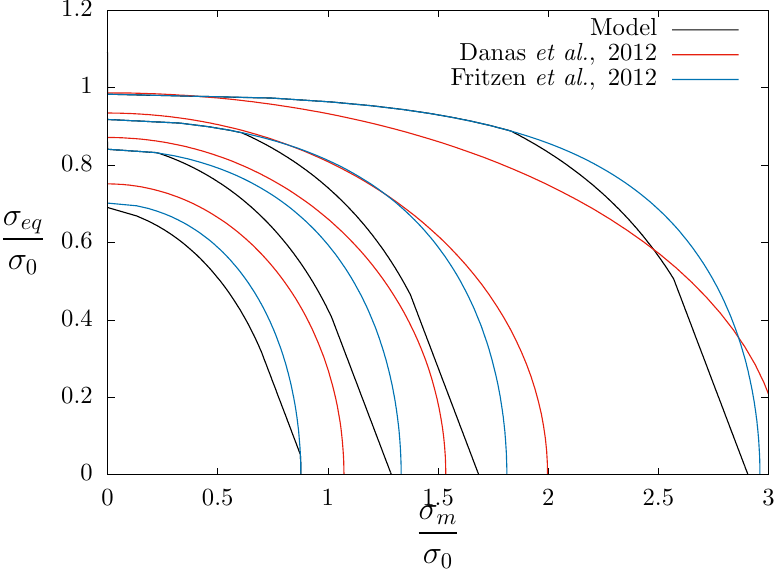}}
\caption{Comparisons of the yield criterion calibrated in this study to the GTN model calibrated in \cite{fritzen2012} and MVAR model \cite{DANAS20122544}, for isotropic materials containing random distributions of spherical voids}
\label{fig10}
\end{figure}

Different models have been proposed in the recent literature to describe the yield surface of isotropic materials containing random distributions of voids. Three of them are compared in Fig.~\ref{fig10}: the first one is the GTN model calibrated in \cite{fritzen2012}, the second one is the model proposed by Danas \textit{et al.} \cite{DANAS20122544}, known as MVAR and based on the variational approach, while the third one corresponds to the model used in this study that combines Fritzen \textit{et al.} GTN model and the inhomogeneous yielding model proposed in \cite{KERALAVARMA2017100,REDDI2019190}. For low porosity - $f = 0.01$ - all models are rather close but significant differences appear when porosity increases. As the model proposed in this study has been calibrated on numerical data, and therefore is a good approximation, Fig.~\ref{fig10} shows that both Fritzen and Danas models overestimate the yield surface, especially for high stress triaxiality as inhomogeneous yielding is not accounted for in these models.

\begin{figure}[H]
  \centering
  \subfigure[]{\includegraphics[height = 5cm]{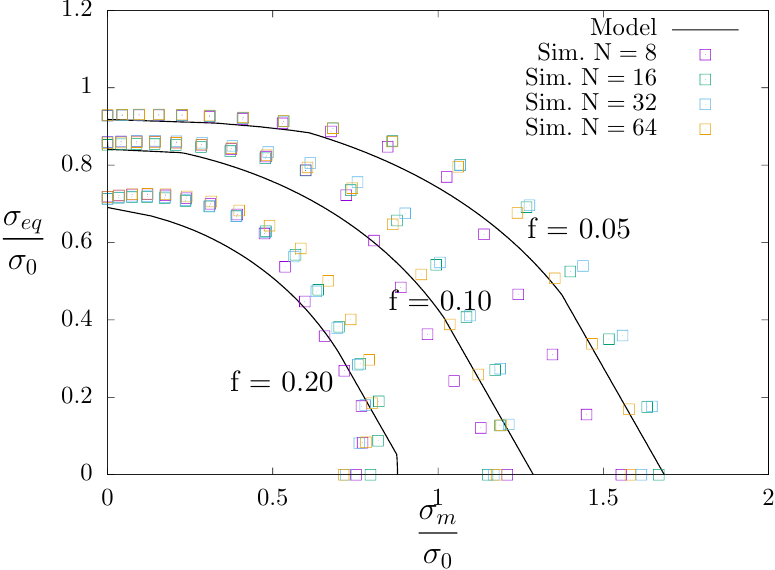}}
  \hspace{0.5cm}
  \subfigure[]{\includegraphics[height = 5cm]{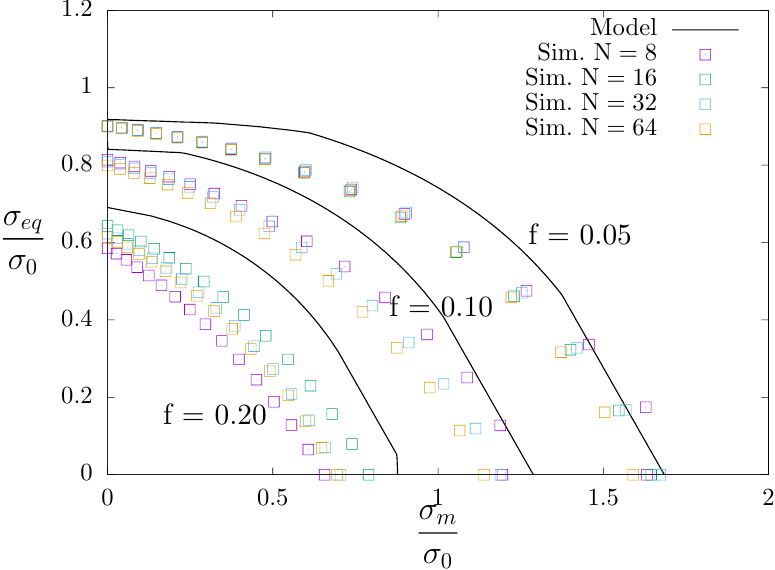}}
\caption{Yield surfaces in the plane mean stress $\sigma_m$ - von Mises equivalent stress $\sigma_{eq}$ determined for different porosities through numerical limit analysis for spheroidal voids with aspect ratio (a) $W = 2$ and (b) $W = 0.5$}
\label{fig11}
\end{figure}

As shown on Fig.~\ref{fig9}, the comparisons between unit cells simulations and predictions of the homogenized model show a satisfactory agreement for three different stress triaxialities, corresponding to an overall evolution of the initially spherical voids from prolate (for $T = 1$) to oblate (for $T = 3$), despite the fact the homogenized model is based on spherical voids. Nevertheless, the effect of void shape on yield surface has been studied numerically, and the results are shown in Fig.~\ref{fig11} for initially spheroidal voids, for aspect ratio $W = 2$ in Fig.~\ref{fig11}a and $W = 0.5$ in Fig.~\ref{fig11}b. In both cases, the assumption of isotropy used in \cite{KERALAVARMA2017100,REDDI2019190} breaks down due to the anisotropy induced by void shape. However, the yield criterion calibrated for $W = 1$ is still in fair agreement with the numerical data for prolate voids Fig.~\ref{fig11}a. The agreement is less good for oblate voids Fig.~\ref{fig11}b, which was also somehow expected as inhomogeneous yielding for oblate voids requires special treatments. Anyhow, the results shown on Fig.~\ref{fig11} call for the extension of the model proposed in \cite{KERALAVARMA2017100,REDDI2019190} to handle void shape effects.

\textcolor{black}{The validation and calibration of the homogenized yield criteria for random porous isotropic materials performed in this study do not, as such, give insights about the domain of applicability of these models. The concept of Representative Volume Element is relevant when its size is small enough compared to the typical size of the structure to be modelled, as well as large enough compared to the size of the microstructure \cite{GLUGE201291}. For random porous materials studied here, the size of the microstructure is set by void radius and inter-void distance. For clustered void distributions observed experimentally \cite{HANNARD2017121}, inter-cluster distance is expected to set the size of the microstructure. These two cases correspond to two potentially different fracture scenarios as proposed in \cite{tekogluphil}. Clearly, the models described in this paper are not relevant to model materials containing voids clusters, and additional simulations are required. Note that these simulations would involve large unit cells that should contain several clusters, each of them involving several voids, which is expected to be computationally very expensive. An alternative strategy would be to perform the numerical homogenization by replacing locally the explicit description of the void clusters by the homogenized models for random porous materials. Besides, the domain of applicability of the models described still need to be determined, especially in situations involving the propagation of a crack. Numerical simulations accounting explicitly for voids in the vicinity of a crack tip \cite{tvergaard2002,hutter2013} have firstly shown different damage growth mechanism - void-by-void \textit{vs.} multiple voids - depending on porosity $f$. Such viewpoint has been challenged recently in \cite{LIU201921} where no such clear distinction between damage mechanisms has been observed on inclusion-driven simulations. Two lengthscales appear in a ductile crack propagation problem: the first one corresponds to crack tip blunting and sets the typical size of the process zone $r_p \sim J/\sigma_0$, where $J$ is the J-integral and measures the intensity of the mechanical loading, and $\sigma_0$ is the yield stress. The second lengthscale is related to the intervoid distance $X_0 \sim R f^{-1/3}$, with $R$ the void radius. Both lengthscales are a priori coupled through the damage process, but set two different regimes: for $r_p \ll X_0$, details about void microstructures become important, and homogenization is not relevant in principle. On the contrary, for $r_p \gg X_0$, many voids are present in the process zone, and homogenized models corresponding to a RVE of randomly distributed voids should be used. In any case, more simulations involving the explicit description of voids appear to be necessary to validate the use of homogenized models.}

\section{Conclusions}

The numerical simulations presented in this paper allow, for isotropic materials containing random distributions of voids, to support the existence of a RVE for inhomogeneous yielding and to provide a calibration of a multi-surface yield criterion accounting for both homogeneous and inhomogeneous yielding. Finite strain simulations, seldom reported in the literature, have further confirmed the existence of the RVE up to the onset of coalescence. The RVE relevance however disappears when inhomogeneous yielding sets in. A homogenized model based on the calibrated yield surfaces leads to a satisfactory agreement with the numerical results up to the onset of coalescence. One perspective of further studies has been described in the discussion section, related to the effect of void shapes for inhomogeneous yielding of porous materials containing random distributions of voids, where anisotropy is induced by void shape. Although such effect is not expected to be predominant for the axisymmetric loading conditions considered in this study - as shown in Figs.~\ref{fig10} and \ref{fig11}, it might become relevant for loading involving shearing. Another perspective, as already outlined in the introduction, is to define when such model should be used in ductile fracture simulations instead of homogenized models based on single voids, especially for inhomogeneous yielding and related to the void-to-void \textit{vs.} multiple voids crack growth scenarii.\\

\noindent
\textbf{Acknowledgements}
The author would like to thank K. Danas for discussions that have triggered part of this study, and L. G\'el\'ebart for the development of the FFT solver \texttt{AMITEX\_FFTP}. Fruitful discussions with J.M. Scherer are also acknowledged.

\newpage

\bibliographystyle{elsarticle-num.bst}
\bibliography{spebib2}

\end{document}